\documentclass{aa}

\usepackage{graphicx}
\usepackage{txfonts}
\usepackage{longtable}
\usepackage{longtable,lscape}
\begin{document}

\title{Chemical analysis of giant stars in the young open cluster NGC~3114 
\thanks{Based on observations made with the 2.2m telescope at the European
Southern Observatory (La Silla, Chile).}}

\author{O. J. Katime Santrich\inst{1}, C. B. Pereira\inst{1} \& N. A. Drake\inst{1,2}}

\offprints{C.B. Pereira}

\institute{Observat\'orio Nacional, Rua Jos\'e Cristino 77, CEP 20921-400, S\~ao Crist\'ov\~ao
Rio de Janeiro, RJ, Brazil.
 \and Sobolev Astronomical Institute, St.~Petersburg State University, 
Universitetski pr.~28, St.~Petersburg 198504, Russia.\\
    \email{osantrich,claudio,drake}@on.br}

  \date{Received; accepted }

\abstract
{Open clusters are very useful targets for examining possible trends in 
galactocentric distance and age, especially when young and old open clusters 
are compared.}
{We carried out a detailed spectroscopic analysis to derive the chemical 
composition of seven red giants in the young open cluster NGC~3114.  
Abundances of C, N, O, Li, Na, Mg, Al, Ca, Si, Ti, Ni, Cr, Y, 
Zr, La, Ce, and Nd were obtained, as well as the carbon isotopic ratio.}
{The atmospheric parameters of the studied stars and their chemical abundances 
were determined using high-resolution optical spectroscopy. We employed the 
local-thermodynamic-equilibrium model atmospheres of Kurucz and the
spectral analysis code {\sc moog}. The abundances of the light elements were 
derived using the spectral synthesis technique.}
{We found that NGC~3114 has a mean metallicity of [Fe/H] = $-$0.01$\pm$0.03. 
The isochrone fit yielded a turn-off mass of $4.2\,M_{\odot}$. The [N/C] ratio 
is in good agreement with the models predicted by first dredge-up. We found 
that two stars, HD~87479 and HD~304864, have high rotational velocities of 15.0~km\,s$^{-1}$ 
and 11.0~km\,s$^{-1}$; HD~87526 is a halo star and is not 
a member of NGC~3114.}
{The carbon and nitrogen abundance in NGC 3114 agree with the field and cluster
giants. The oxygen abundance in NGC~3114 is lower compared to the field giants.
The [O/Fe] ratio is similar to the giants in young clusters.
We detected sodium enrichment in the analyzed cluster giants. 
As far as the other elements are concerned, their [X/Fe] ratios follow
the same trend seen in giants with the same metallicity.}

\keywords{Galaxy: open clusters and associations: individual : {NGC~3114} --- stars: abundances --- stars: fundamental parameters --- 
stars: individual : {HD~87109, HD~87479, HD~87526, HD~87566, HD~87833, HD~304859, HD~304864}}
\authorrunning{O. J. Katime Santrich et al.}
\titlerunning{Young open cluster NGC~3114}
\maketitle

\newpage

\section{Introduction}

\par Open clusters are very useful targets for probing the chemical
evolution of the Galaxy in addition to H\,{\sc ii} regions,
planetary nebulae, cepheid variables, and OB stars.  The chemical
abundances of the open clusters at several galactocentric distances
help us to investigate whether an abundance gradient in the galactic
disk exists and so better understand the structure and evolution of
the Milky Way.  The advantage of open clusters is that they have a
wide range of age and distance, in other words, they have formed at
several epochs and at different galactocentric distances. Because of
the range in age, from approximately 0.1 up to 12.0 Gyr, we can
investigate whether there are any noticeable difference in the abundance
ratios [X/Fe] due to evolutionary effects when young and old
clusters are compared. Because young clusters are composed of more
massive stars than old clusters, a comparison of giants of different
clusters would show differences in abundances (if any) due to some
nuclear process that might have occurred in these more massive
stars. In addition, we could also compare how processes such as the
first dredge-up and extra-mixing affect the chemical composition of
these higher mass giant stars.

\par In this work we will analyze seven giants in the young open
cluster NGC~3114 using high-resolution spectroscopy with the aim of
obtaining their abundance pattern.  Some previous photometric
studies have been done for NGC~3114 (Jankowitz \& McCosh 1963; Schneider \&
Weiss 1988; Clari\'a et al. 1989; Sagar \& Sharpless 1991; Carraro \&
Patat 2001; Paunzen et al. 2003), but a high-resolution spectroscopic
analysis of the giants in this cluster has not been done yet. As we
shall see, the giants in NGC~3114, like many other giants in young
clusters, have a lower oxygen abundance compared to the giants in old
clusters. We also detected a sodium enrichment in the atmospheres of
the giants in this cluster.  Here we have considered as young clusters
those with ages less than 1.0 Gyr, and as old those with ages higher
than 1.0 Gyr. The abundances of other elements, the lighter ones
(carbon, nitrogen, and lithium) and the heavier (aluminum to neodymium), 
as well as the carbon isotopic ratio were also obtained.

\par The open cluster NGC~3114 (C\,1001-598, VDBH~86; $\alpha=10^{h} 02^{m}$.0,
$\delta=-60^\circ06^{'}$(2000.0); $l=283^\circ$, $b= -04^\circ$) is 
projected onto the Carina complex (Carraro \& Patat 2001).  
These authors presented a CCD photometric study of NGC~3114
and obtained $UBVRI$ magnitudes down to $V\approx22.0$. Their results
for the age, the distance, and color excess are 1.6$\times$10$^{8}$ years, 
920$\pm$50 pc, and $E(B-V)=0.07$.  
Similar values were also found by Gonz\'alez \& Lapasset (2001).  Because
NGC~3114 lies in a region of heavy contamination by field stars, it is
difficult to distinguish between members and non-members of NGC~3114.
Thanks to the radial-velocity survey of Gonz\'alez \& Lapasset (2001)
and more recently of Mermilliod et al. (2008), several stars were
identified as members of this cluster, including the seven red giants
analyzed in this work.  The CMD of NGC~3114 is shown in Fig.~1, with
our sample stars (red squares) for identification. From the isochrone
fit, we derived a turn-off mass of $4.2\,M_{\odot}$. The age and turn-off 
mass of NGC~3114 are similar to the young cluster M\,11 studied by Gonzalez
 \& Wallerstein (2000).

\begin{figure}
\centering
\includegraphics[width=9.5cm]{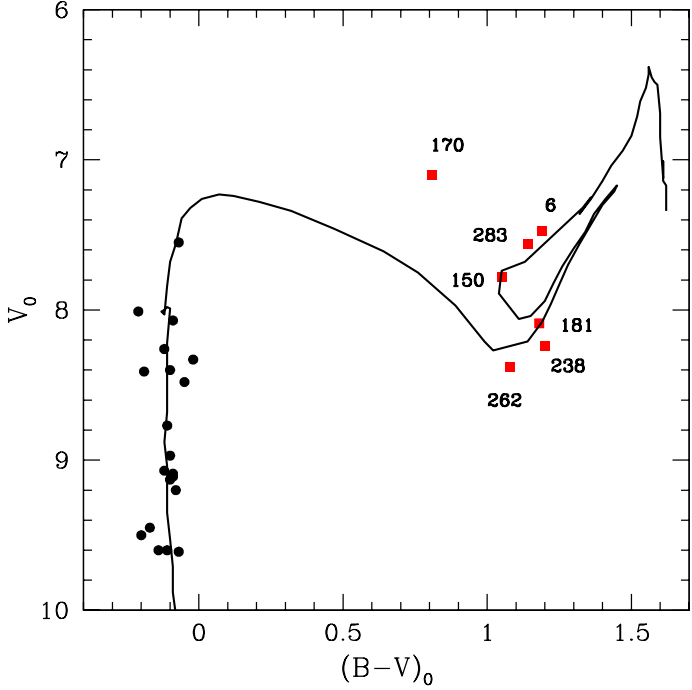}
\caption{Reddening corrected color-magnitude diagram of NGC~3114.
Black filled circles are kinematically main-sequence member stars 
from Tables~2 and 3 of Gonz\'alez \& Lapasset (2001). Our program 
stars are identified by red filled squares. We also show the isochrone
of Bertelli et al. (1994) for $\log t$\,=\,8.2 (0.16 Gyr), Z\,=\,0.02, and Y\,=\,0.28.}
\end{figure}

\section{Observations}

\par The high-resolution spectra of the stars analyzed in this work were
obtained at the 2.2m ESO telescope at La Silla (Chile) with the FEROS
(Fiberfed Extended Range Optical Spectrograph) echelle spectrograph (Kaufer et
al. 1999).  The FEROS spectral resolving power is $R$=48\,000, corresponding to
2.2 pixels of 15$\mu$m, and the wavelength coverage is from 3\,800\,\AA\, to
9\,200\,\AA.  The stars were selected from the radial velocity survey of
Mermilliod et al. (2008).  Table~1 gives the log of
observations and additional information about the observed stars. The nominal
S/N ratio was evaluated by measuring the rms flux fluctuation in selected
continuum windows, and the typical values were S/N = 100 -- 150.

\begin{table*}
\caption{Log of the observations and relevant information for the target stars.
ID and HD numbers, $V$, $B-V$, and radial velocities were taken from Mermilliod et al. (2008) 
(Cols.~1 -- 5). Our values for the radial and rotational velocities are given in 
Cols.~6 and 7. The last three columns provide the dates of observation, 
exposure times, and spectral types.}
\label{tab:logtable}
\begin{tabular}{lccccccccc}
\hline
ID  &  HD   & $V$ & $B-V$&   RV    &     RV$^{a}$   &  $v\sin i$     &  Date obs  &   Exp   &  SpT$^{c}$      \\
    &       &     &      & km\,s$^{-1}$ &  km\,s$^{-1}$  &  km\,s$^{-1}$ &       &  sec &     \\\hline
6   & 87109 & 7.6 & 1.29 & $-$1.43$\pm$0.23 & $-$1.31$\pm$0.49$^{b}$ & 6.5$\pm$1.0  & 2008 Apr 9  & 600 & G9\,II  \\
150 & 87479 & 7.9 & 1.17 & $-$2.19$\pm$0.51 & $-$1.71$\pm$0.61      & 15.0$\pm$2.0 & 2008 Apr 9  & 420 & G8\,II-III  \\
170 & 87526 & 7.3 & 0.89 & $-$1.95$\pm$0.40 & $-$2.38$\pm$0.17      & 8.0$\pm$1.0  & 2008 Apr 9  & 600 & G1\,II-III \\
181 & 87566 & 8.3 & 1.28 & $-$2.18$\pm$0.10 & $-$2.27$\pm$0.23      & $<$ 4.5      & 2008 Apr 10 & 420 & K1\,III  \\
238 & 304859& 8.5 & 1.27 & $-$1.72$\pm$0.18 & $-$1.40$\pm$0.34      & $<$ 4.5      & 2008 Dec 22 & 600 & K1\,III  \\
262 & 87833 & 8.6 & 1.16 & $-$1.20$\pm$0.23 & $-$1.22$\pm$0.17      & 8.0$\pm$1.0  & 2008 Apr 10 & 600 & G9\,III  \\
283 & 304864& 7.7 & 1.25 & $-$1.73$\pm$0.31 & $-$1.41$\pm$0.36      & 11.0$\pm$2.0 & 2008 Dec 22 & 600 & G9\,II-III  \\
\hline
\end{tabular}
\tablefoot{\\
\tablefoottext{a}{This work}\\
\tablefoottext{b}{$-$1.33~km\,s$^{-1}$ (de Medeiros et al. 2002)}\\
\tablefoottext{c}{Gonz\'alez \& Lapasset (2001)}\\
}
\end{table*}

\section{Analysis and results}

\subsection{Line selection, measurement and oscillator strengths}

\par The spectra of the stars show many atomic absorption lines of Fe\,{\sc
  i} and Fe\,{\sc ii} as well as transitions due to Na\,{\sc i}, Mg\,{\sc i},
Al\,{\sc i}, Si\,{\sc i}, Ca\,{\sc i}, Ti\,{\sc i}, Cr\,{\sc i}, Ni\,{\sc i},
Y\,{\sc ii}, Zr\,{\sc i}, La\,{\sc ii}, Ce\,{\sc ii}, and Nd\,{\sc ii}. We
have chosen a set of lines sufficiently unblended to yield reliable
abundances. In Table~2 we list the Fe\,{\sc i} and
Fe\,{\sc ii} lines employed in the analysis, the lower excitation
potential of the transitions, $\chi$ (ev), the $gf$-values, and the measured
equivalent widths. The equivalent widths were obtained by fitting Gaussian
profiles to the observed ones. The $gf$-values for the Fe\,{\sc i} and
Fe\,{\sc ii} lines in Table~2 were taken from Lambert et al.
(1996) and Castro et al. (1997).\addtocounter{table}{1}

\par To avoid systematic errors in the abundance
  analysis we determined the solar abundance using the same program
  and line lists. We used the solar spectrum of Ganimede
  taken with the ESO spectrograph HARPS. We used the solar atmosphere
  model of Kurucz (1993) {\sl T}$_{\rm eff}$\,=\,5\,777, $\log
  g$\,=\,4.44 and the microturbelent velocity of $\xi$=0.75
  km\,s$^{-1}$ (Pavlenko et al. 2012).  The results of our analysis
  are given in Table 3 and are in good agreement with those given by
  Grevesse \& Sauval (1998) (hereafter GS98) which are also shown in
  Table 3. We derived a low aluminum abundance and adopted the
  abundance of GS98. A low solar aluminum abundance was also found by
  Reddy et al. (2012). According to Pancino et al. (2010) the lines of
  aluminum at 6696 \AA\, and 6698 \AA\, give lower abundances than
  other aluminum lines. Carbon, nitrogen, and oxygen abundances were
  determined using the spectrum synthesis technique. The abundances of
  carbon and nitrogen were determined using the lines of the CN and
  C$_2$ molecules. We assembled the line lists and they are the same
  as in Drake \& Pereira (2008, 2011) and Pereira \& Drake
  (2009) who studied the chemically peculiar metal-poor stars
  HD~104340, HD~206983, HD~10613, BD+04$\degr$2466, and the Feh-Duf
  star.  For the determination of the oxygen abundance we used the
  [O\,{\sc i}] forbidden line at 6300.304~\AA.  The derived [X/H] and
  [X/Fe] ratios in this work were calculated using the solar abudances
  given in Table 3.

\begin{table}
\caption{Our solar abundances compared with the photospheric
abundances of Grevesse \& Sauval (1998) (GS98). The number in parentheses
indicates the number of lines used for abundance analysis. 
For carbon, nitrogen, and oxygen the number in parentheses means the number 
of spectral regions used in the synthesis.}
\begin{tabular}{lcc}\hline
Species         &  $\log\varepsilon_{\odot}$ & $\log\varepsilon_{\odot}$\\\hline
                &                 & GS98\\\hline
C  & 8.62$\pm$0.05(2)  & 8.52$\pm$0.06\\
N  & 7.96$\pm$0.05(1)  & 7.92$\pm$0.06\\
O  & 8.77$\pm$0.08(1)  & 8.83$\pm$0.06\\
Na & 6.34$\pm$0.05(3)  & 6.33$\pm$0.03\\
Mg & 7.60$\pm$0.08(2)  & 7.58$\pm$0.05\\
Al & 6.32$\pm$0.14(2)  & 6.47$\pm$0.07\\
Si & 7.57$\pm$0.01(3)  & 7.55$\pm$0.05\\
Ca & 6.39$\pm$0.09(13) & 6.36$\pm$0.02\\
Ti & 4.99$\pm$0.06(44) & 5.02$\pm$0.06\\
Cr & 5.67$\pm$0.09(29) & 5.67$\pm$0.03\\
Fe\,{\sc i}  & 7.54$\pm$0.08(74) & 7.50$\pm$0.05\\
Fe\,{\sc ii} & 7.54$\pm$0.03(11) &  ---        \\
Ni & 6.31$\pm$0.05(47) &  6.25$\pm$0.04\\
Y  & 2.26$\pm$0.05(3)  &  2.24$\pm$0.03\\
Zr & 2.64$\pm$0.06(3)  &  2.60$\pm$0.02\\
La & 1.22$\pm$0.09(2)  &  1.17$\pm$0.07\\
Ce & 1.58$\pm$0.02(4)  &  1.58$\pm$0.09\\
Nd & 1.60$\pm$0.05(3)  &  1.50$\pm$0.06\\\hline
\end{tabular}
\end{table}

\subsection{Atmospheric parameters}

\par The determination of stellar atmospheric parameters such as
effective temperature ($T_{\rm eff}$), surface gravity ($\log g$),
microturbulence ($\xi$), and metallicity, as given by [Fe/H]
(throughout we use the notation [X/H]=$\log(N_{\rm X}/N_{\rm
  H})_{\star} -\log(N_{\rm X}/N_{\rm H})_{\odot}$) are prerequisites
for the determination of photospheric abundances. The atmospheric
parameters were determined using the local thermodynamic equilibrium
(hereafter LTE) model atmospheres of Kurucz (1993) and the spectral
analysis code {\sc moog} (Sneden 1973).

\par The solution of the excitation equilibrium used to derive the
effective temperature ($T_{\rm eff}$) was defined by a zero slope of
the trend between the iron abundance derived from Fe\,{\sc i} lines
and the excitation potential of the measured lines.  The
microturbulent velocity ($\xi$) was found by constraining the
abundance, determined from individual Fe\,{\sc i} lines, to show no
dependence on $W_{\lambda}/{\lambda}$.  The solution thus found is
unique, depending only on the set of Fe\,{\sc i} and Fe\,{\sc ii} lines and 
the atmospheric model employed. As a by-product this yields the metallicity 
[Fe/H] of the star. The value of $\log g$ was determined by means of the 
ionization balance assuming LTE. The final adopted atmospheric parameters are 
given in Table~4.

\begin{table*}
\caption{Adopted atmospheric parameters and metallicity. For [Fe\,{\sc i}/H] and [Fe\,{\sc ii}/H]
we also show the standard deviation and the number of lines employed .}
\label{tab:atmparam}
\begin{tabular}{lccccc}\hline
Star & $T_{\rm eff}$ & log $g$ & $\xi$ & [Fe\,{\sc i}/H]$\pm$ $\sigma$ (\#) & [Fe\,{\sc ii}/H]$\pm$ $\sigma$ (\#)\\
          &    K        &      & km\,s$^{-1}$  &  &        \\\hline 
HD  87109 &  4\,700 &  1.2  & 2.0 &  $-$0.04$\pm$0.13\,(51)  &  $-$0.06$\pm$0.14\,(9)  \\ 
HD  87479 &  4\,900 &  1.8  & 2.4 &  $-$0.03$\pm$0.16\,(32)  &  $-$0.02$\pm$0.24\,(4)  \\ 
HD  87526 &  5\,300 &  1.5  & 2.9 &  $-$0.75$\pm$0.10\,(71)  &  $-$0.76$\pm$0.09\,(13) \\
HD  87566 &  4\,500 &  1.6  & 1.6 &  $+$0.00$\pm$0.18\,(56)  &  $-$0.01$\pm$0.11\,(11) \\
HD  87833 &  4\,900 &  2.2  & 1.9 &  $-$0.05$\pm$0.15\,(42)  &  $-$0.04$\pm$0.12\,(9)  \\
HD 304859 &  4\,500 &  1.6  & 1.6 &  $+$0.04$\pm$0.18\,(51)  &  $+$0.05$\pm$0.09\,(10) \\
HD 304864 &  4\,700 &  1.2  & 2.2 &  $+$0.01$\pm$0.15\,(30)  &  $+$0.03$\pm$0.22\,(3)  \\\hline 
\end{tabular}
\end{table*}

\par Previous determination of atmospheric parameters in one of the
stars in this cluster, HD~87566 (number 181), was done by Santos et
al. (2009) who found two different solutions according to the line lists used:
(i) $T_{\rm eff}\!=\!4561\pm$45~K, $\log g\! =\! 1.92\!\pm\!0.32$~dex,
$\xi\!=\!2.02\!\pm\! 0.04$~km\,s$^{-1}$, and [Fe/H]\,$\!=\!-0.11\!\pm\!0.13$~dex, 
and (ii) $T_{\rm eff}\!=\!4384\pm$147~K, $\log g$\,=\,1.65$\pm$0.36~dex,
$\xi$\,=\,2.01$\pm$0.15~km\,s$^{-1}$, and [Fe/H]\,$\!=\! -0.23\!\pm\!0.17$~dex.

\par The internal errors in our adopted effective temperatures
($T_{\rm eff}$) and microturbulent velocities ($\xi$) can be
determined from the uncertainty in the slopes of the relationships of
Fe\,{\sc i} abundance $versus$ excitation potential and Fe\,{\sc i}
abundance $versus$ reduced equivalent width ($W_{\lambda}/\lambda$). 
The standard deviation in $\log~g$ was set by
changing this parameter around the adopted solution until the
difference between Fe\,{\sc i} and Fe\,{\sc ii} mean abundance
differed by exactly one standard deviation of the [Fe\,{\sc i}/H] mean
value. Based on the above description, we estimate typical
uncertainties in atmospheric parameters of the order of $\pm$140\,K,
$\pm$0.20 -- 0.25~dex, and $\pm$0.3 km\,s$^{-1}$, respectively, for
$T_{\rm eff}$, $\log g$, and $\xi$.

\par We compare the derived spectroscopic gravities for each star 
of the cluster with those obtained from the equation

\begin{eqnarray}
\log g_{\star}\;  & = & \log \frac{M_{\star}}{M_{\odot}}  
+ 0.4\left(V-A_{ V}+BC\right)  \nonumber \\
& &  {{\,}\atop{\,}} + 4\log T_{\rm eff} - 2 \log r\; ({\rm kpc}) - 16.5.
\end{eqnarray}

For the mass we adopted the turn-off mass of the cluster and for the
distance and color excess we adopted $r\!=\!920$~pc and $E(B-V)=0.07$.
Bolometric corrections were taken from Alonso et al. (1999).  We found
a {\sl mean} difference of 0.27$\pm$0.2 between the spectroscopic and
evolutionary gravities as given from the above equation.  This mean
difference lies in the same range of the uncertainties introduced by
the standard deviation in $\log g$.

\subsection{Abundance Analysis}

\par The abundances of chemical elements were determined with the LTE
model atmosphere techniques.  In brief, equivalent widths are
calculated by integration through a model atmosphere and by comparing
this with the observed equivalent widths. The calculations are repeated,
changing the abundance of the element in question, until a match is
achieved. The current version of the line-synthesis code {\sc moog}
(Sneden 1973) was used to carry out the calculations.  Table~5 shows
the atomic lines used to derive the abundances of the elements. Atomic
parameters for several transitions of Ti, Cr, and Ni were retrieved
from the National Institute of Science and Technology
Atomic Spectra Database (Martin 2002).\addtocounter{table}{1}

\par The carbon, nitrogen, and oxygen abundances were
derived as described in Sect.~3.1. The abundances of the
CNO elements are interdependent because of the association of carbon and
oxygen in CO molecules in the atmospheres of cool giants, therefore
the CNO abundance determination procedure was iterated until all the
abundances of these three elements agreed.  The $^{12}$C/$^{13}$C
isotopic ratios were determined using the lines of the CN
molecules.  In Figs. 2 and 3 we show the observed and synthetic
spectra in the regions of the [O\,{\sc i}] $\lambda$6300~\AA\, line
and $\sim\!\lambda$8000~\AA\, containing the $^{12}$CN and the
$^{13}$CN molecular lines.

\begin{figure*} 
   \centering
   \includegraphics[width=9.5cm]{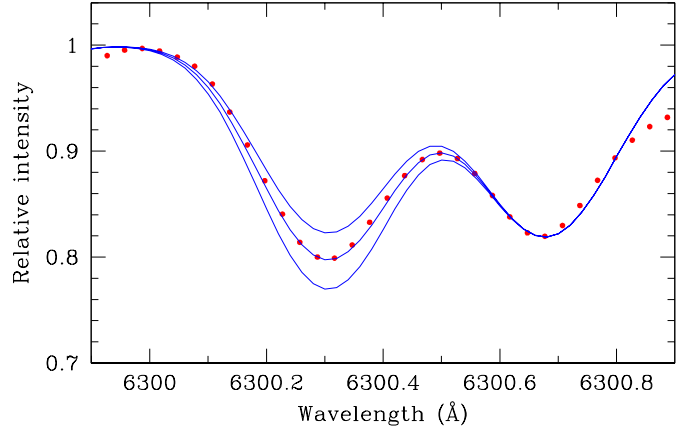}
   \caption{Observed (dotted red line) and synthetic (solid blue line)
spectra of HD 87109 in the region around the oxygen forbidden line at $\lambda$6300 \AA.
The synthetic spectra were calculated with the oxygen 
abundances of [O/Fe] = $-$0.37, $-$0.27, and $-$0.17.}
\end{figure*}

\begin{figure*} 
   \centering
   \includegraphics[width=9.5cm]{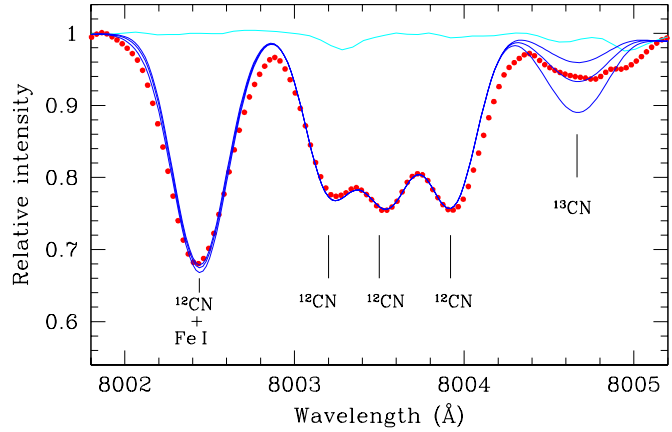}
   \caption{Observed (dotted red line) and synthetic (solid blue line) spectra of the star HD 87109 in the region containing 
the $^{12}$CN and the $^{13}$CN molecular lines. In the synthetic spectra we show the synthesis for three values of
$^{12}$C/$^{13}$C ratios, 12.0, 21.0, and 36.0. The light blue solid line shows the spectrum of the hot star used to map the 
telluric water lines.}
\end{figure*}

\par Lithium abundance was derived from the synthetic spectra matches
to the Li\,{\sc i} $\lambda 6708$~\AA~resonance doublet. The CN
lines in the vicinity of the Li\,{\sc i} doublet were included in the
line list.  The wavelengths and oscillator strengths for the
individual hyperfine and isotopic components of the lithium lines were
taken from Smith et al.  (1998) and Hobbs et al. (1999).  A solar
$^6$Li/$^7$Li isotopic ratio ($^6$Li/$^7$Li$=0.081$) was adopted in
the calculations of the synthetic spectrum.

\par Table~6 shows the abundance of light elements (lithium, carbon,
nitrogen, and oxygen). Except for lithium, we show the abundances for each star 
in the notation [X/Fe]. In Table~6 we also provide the
$^{12}$C/$^{13}$C isotopic ratio and show the mean cluster abundances
and their standard deviations, i.e., the scatter around the mean.  For
the calculation of the mean abundance of the cluster we do not take
into account the results for HD~87526 which is a non-member of the
cluster (Sect.~4.2).  Tables~7 and 8 provide, for each star, the
derived mean abundances of the elements in the notation [X/Fe] and the
mean cluster abundances with their respective standard
deviations. Furthermore, we obtained the abundances of the heavy elements
created by slow neutron capture reactions (s-process): Y, Zr, La, Ce,
and Nd.  The {\sl mean} abundances of these heavy elements, s in
the notation [s/Fe], are included in Table 8.  We do not measure the
barium abundance in our stars because all barium lines have equivalent
widths higher than 200~m\AA\, and therefore they will not lie at the
linear part of the curve of growth (Hill et al. 1995). However, since
we have measured several lines of other elements synthesized by the
s-process, we think that we probe this nucleosynthesis process in our
cluster giants fairly well.

\begin{table*}
\caption{Light element abundances. The lithium abundance is given in the notation 
$\log\varepsilon$. For the carbon, oxygen, and nitrogen we give abundance ratios ([X/Fe]). 
The last column provides the $^{12}$C/$^{13}$C isotopic ratio.}
\begin{tabular}{lccccc}
\hline
Star     &$\log\varepsilon$(Li)& [C/Fe] & [N/Fe] & [O/Fe] & $^{12}$C/$^{13}$C \\\hline
HD 87109 &   0.9    &  $-$0.29 & $+$0.34$\pm$0.04(8) & $-$0.20 & 21 \\
HD 87479 &   1.3    &  $-$0.27 & $+$0.54$\pm$0.05(8) & $-$0.11 & $\geq$20 \\
HD 87526 &   1.1    &  $+$0.33 & $+$1.13$\pm$0.08(6) &  $+$0.38 & $\geq$24 \\
HD 87566 &   0.3    &  $-$0.23 & $+$0.43$\pm$0.06(8) & $-$0.07 & 11 \\
HD 87833 &   1.3    &  $-$0.35 & $+$0.40$\pm$0.04(9) & $-$0.02 & 24\\
HD 304859 &  0.3    &  $-$0.37 & $+$0.17$\pm$0.05(8) & $-$0.21 & 16 \\
HD 304864 &  1.2    &  $-$0.43 & $+$0.35$\pm$0.04(9) & $-$0.32 & $\geq$16 \\\hline
mean$^a$  &  0.9$\pm$0.4 & $-$0.31$\pm$0.07 & $+$0.38$\pm$0.12 & $-$0.16$\pm$0.10 & $\geq$18 \\\hline
\end{tabular}
\tablefoot{Results for HD~87526 were excluded from the obtained mean abundance. \\}
\end{table*}

\begin{table*}
\caption{Mean abundance ratios ([X/Fe]) for the elements from Na to Cr.
We also provide the number of lines used for the abundance determination and the abundance 
dispersion among the lines of those elements with more than three available
lines.}
\begin{tabular}{lccccccc}
\hline
Star     & [Na/Fe](NLTE)$^a$ & [Mg/Fe] & [Al/Fe] & [Si/Fe] & [Ca/Fe] & [Ti/Fe] & [Cr/Fe] \\\hline

HD 87109 & $+$0.36(2) & $-$0.06$\pm$0.19(5)  & $-$0.05$\pm$0.18(4) & $+$0.03$\pm$0.18(9) 
& $-$0.01$\pm$0.16(3) & $-$0.11$\pm$0.20(14) & $-$0.01$\pm$0.26(20) \\ 

HD 87479 & $+$0.26(1)  & --- & $+$0.13$\pm$0.21(4) & $+$0.05(2) & $-$0.02(2) 
& $+$0.06$\pm$0.19(9)  & $-$0.07$\pm$0.21(4) \\ 

HD 87526 & $+$0.42(2) & $+$0.50$\pm$0.25(3) & $+$0.29$\pm$0.26 & $+$0.35$\pm$0.17(8)
& $+$0.11$\pm$0.12(7) & $-$0.07$\pm$0.14(13) & $+$0.19$\pm$0.14(8) \\ 

HD 87566 & $+$0.26(2) & $+$0.05$\pm$0.25(8)  & $-$0.16$\pm$0.19(5) & $-$0.01$\pm$0.21(9) 
& $-$0.02$\pm$0.23(3) & $-$0.12$\pm$0.21(16) & $-$0.05$\pm$0.28(17) \\ 

HD 87833 & $+$0.36(2) & $+$0.01$\pm$0.22(8) & $+$0.08$\pm$0.25(5) & $+$0.04$\pm$0.18(6)
& $-$0.11$\pm$0.21(5) & $+$0.06$\pm$0.21(17) & $+$0.05$\pm$0.24(16) \\ 

HD 304859 & $+$0.19(2) & $+$0.12$\pm$0.30(7) & $-$0.11$\pm$0.20(6) & $+$0.13$\pm$0.20(8) 
& $+$0.01$\pm$0.32(3)  & $-$0.14$\pm$0.21(16) & $-$0.05$\pm$0.25(19) \\  

HD 304864 & $+$0.19(2) & --- & --- & $+$0.18$\pm$0.19(6) & $-$0.13(2) 
& $-$0.13$\pm$0.23(10) & $-$0.02$\pm$0.21(8) \\\hline  

mean$^b$ & $+$0.27$\pm$0.08 & $+$0.03$\pm$0.08 & $-$0.03$\pm$0.10 & $+$0.07$\pm$0.07
& $-$0.05$\pm$0.05 & $-$0.06$\pm$0.06 & $-$0.03$\pm$0.02 \\  
\hline
\end{tabular}
\tablefoot{\\
\tablefoottext{a}{[Na/Fe] accounts for the NLTE effects calculated as in Gratton et al. (1999), see text.}\\
\tablefoottext{b}{Abundance ratios of HD~87526 were excluded from the obtained mean.}\\}
\end{table*}

\begin{table*} 
\caption{Mean abundance ratios ([X/Fe]) for Ni and the heavy elements.
We also provide the number of lines used for the abundance determination and
the abundance dispersion among the lines for those elements with more than
three available lines. The last column of the table shows the mean s-process abundance.}
\begin{tabular}{lccccccc}
\hline
Star      &  [Ni/Fe] & [Y/Fe] & [Zr/Fe] & [La/Fe] & [Ce/Fe] & [Nd/Fe] & [s/Fe]  \\\hline 

HD 87109  & $-$0.08$\pm$0.22(36) & $+$0.09(2) & $-$0.05$\pm$0.21(7) & $+$0.19$\pm$0.16(4)  
& $+$0.10$\pm$0.18(5) & $+$0.07$\pm$0.20(7) & $+$0.08$\pm$0.06 \\ 

HD 87479  & $-$0.15$\pm$0.21(16) & $-$0.01(2) &  --- &  ---  &  --- &  --- & --- \\ 

HD 87526  & $+$0.01$\pm$0.16(21) & $-$0.07$\pm$0.21(5) &  --- &  $-$0.02$\pm$0.21(4)
&  $+$0.16$\pm$0.17(12) &  $+$0.17$\pm$0.19(11) & $+$0.06$\pm$0.09 \\ 

HD 87566  & $+$0.01$\pm$0.28(33) & $-$0.03$\pm$0.23(4) & $-$0.18$\pm$0.19(9) & $+$0.12(2)
& $+$0.13$\pm$0.21(6) & $+$0.08$\pm$0.22(4) & $+$0.02$\pm$0.11 \\ 

HD 87833  & $+$0.06$\pm$0.19(33) & $+$0.01$\pm$0.19(3) & $+$0.17$\pm$0.16(6) & $+$0.24(2)
& $+$0.08$\pm$0.16(4) & $+$0.22$\pm$0.19(10) & $+$0.14$\pm$0.10 \\ 

HD 304859 & $+$0.02$\pm$0.24(37) & $+$0.06$\pm$0.24(4) & $-$0.12$\pm$0.24(11) & $+$0.17$\pm$0.19(4) 
& $+$0.10$\pm$0.20(6) & $+$0.04$\pm$0.25(6) & $+$0.05$\pm$0.07 \\ 

HD 304864 & $-$0.21$\pm$0.22(26) & $-$0.01(1) & $-$0.13$\pm$0.19(3) & $-$0.01(2) & ---
& $+$0.07$\pm$0.19(6) & $-$0.02$\pm$0.07 \\\hline 

mean$^a$ & $-$0.06$\pm$0.08 & $+$0.02$\pm$0.04 & $-$0.06$\pm$0.09 & $+$0.14$\pm$0.09 
& $+$0.10$\pm$0.02 & $+$0.10$\pm$0.07 & $+$0.05$\pm$0.06 \\
\hline
\end{tabular}
\tablefoot{
\tablefoottext{a}{Abundance ratios of HD~87526 were excluded from the obtained mean.}\\
}
\end{table*}

\subsection{Abundance uncertainties}

\par The uncertainties in the derived abundances for the program stars
are dominated by three main sources: the stellar parameters, the
measured equivalent widths, and the $gf$-values. The errors in the
$gf$-values were discussed by Smith, Cunha \& Lambert (1995) and we
thus refer to this paper for a detailed discussion.

\par The abundance uncertainties due to the errors in the stellar
atmospheric parameters $T_{\rm eff}$, $\log$~g, and $\xi$ were
estimated by changing these parameters by their standard errors and
then computing the changes incurred in the element abundances.  This
technique has been applied to the abundances determined from
equivalent line widths as well as to those determined via spectrum
synthesis. The results of these calculations for HD 87109 are
displayed in Cols.~2 to 6 of Table 9.  The abundance variations for
the other stars show similar values.

\par The abundance uncertainties due to the errors in the equivalent
width measurements were computed from the expression provided by
Cayrel (1988). The errors in the equivalent widths are set,
essentially, by the signal-to-noise ratio and the resolution of the
spectra. In our case, having a resolution of 48\,000 and a typical S/N
ratio of 150, the expected uncertainties in the equivalent widths are
about 2--3 m{\AA}. For all measured equivalent widths, these
uncertainties lead to lower errors in the abundances than those coming
from the uncertainties in the stellar parameters.

\par Under the assumption that the errors are independent, they can be 
combined quadratically so that the total uncertainty is given by \[\sigma = \sqrt{\sum_{i=1}^{N} \sigma^{2}_{i}}\, . \]
These final uncertainties are given in the Col.~7 of Table 9. The last column 
gives the abundance dispersion observed among the lines for those elements with 
more than three available lines.

\par Table 9 shows that neutral elements are more sensitive to the
 variations in the temperature while singly-ionized elements are more
 sensitive to the variations in $\log$~g. For the elements whose
 abundance is based on stronger lines, such as the lines of calcium
 and yttrium, the error introduced by the microturbulence is significant.

\begin{table*}
\caption{Abundance uncertainties for HD~87109. The second
column gives the variation of the abundance caused by the variation in
$T_{\rm eff}$. The other columns refer to the variations in the abundances caused by
 $\log g$, $\xi$, [Fe/H], and $W_\lambda$. The seventh column gives
the compounded rms uncertainty of the second to sixth column. The last column gives the 
abundance dispersion observed among the lines for those elements with more than three available
lines.}
\centering
\begin{tabular}{lccccccc}\hline\hline
Species & $\Delta T_{\rm eff}$ & $\Delta\log g$ & $\Delta\xi$ & $\Delta$[Fe/H] & $\Delta W_{\lambda}$
& $\left( \sum \sigma^2 \right)^{1/2}$ & $\sigma_{\rm obs}$\\
$_{\rule{0pt}{8pt}}$ & $+140$~K & $+0.4$ & $+$0.3 km\,s$^{-1}$ & 0.1 & $+$3 m\AA &  \\
\hline     
Li\,{\sc i} &  $+$0.20 &   0.00 &  $-$0.01  &  $-$0.01 & --- & 0.20 & --- \\  
Fe\,{\sc i}    & $+$0.12  & $+$0.02 & $-$0.14 & $-$0.01 & $+$0.04 & 0.19 & 0.13 \\
Fe\,{\sc ii}   & $-$0.12  & $+$0.18 & $-$0.18 & $-$0.02 & $+$0.04 & 0.28 & 0.14 \\
Na\,{\sc i}    & $+$0.13  & $-$0.02 & $-$0.16 & $-$0.01 & $+$0.04 & 0.21 & --- \\
Mg\,{\sc i}    & $+$0.07  & $-$0.01 & $-$0.05 & $-0$.01 & $+$0.04 & 0.10 & 0.14 \\
Al\,{\sc i}    & $+$0.09  & $-$0.01 & $-$0.06 &    0.00 & $+$0.04 & 0.12 & 0.13 \\
Si\,{\sc i}    & $-$0.01  & $+$0.07 & $-$0.08 & $+$0.01 & $+$0.04 & 0.11 & 0.13 \\
Ca\,{\sc i}    & $+$0.17  & $-$0.02 & $-$0.18 & $-$0.02 & $+$0.04 & 0.25 & 0.10 \\
Ti\,{\sc i}    & $+$0.23  & $-$0.01 & $-$0.13 & $-$0.02 & $+$0.05 & 0.27 & 0.15 \\
Cr\,{\sc i}    & $+$0.16  & $-$0.01 & $-$0.08 & $-$0.01 & $+$0.05 & 0.19 & 0.22 \\
Ni\,{\sc i}    & $+$0.10  & $+$0.06 & $-$0.11 & $+$0.02 & $+$0.05 & 0.17 & 0.18 \\
Y\,{\sc ii}    &    0.00  & $+$0.17 & $-$0.20 & $+$0.04 & $+$0.06 & 0.27 & 0.08 \\
Zr\,{\sc i}    & $+$0.26  & $-$0.01 & $-$0.02 & $-$0.02 & $+$0.06 & 0.27 & 0.16 \\
La\,{\sc ii}   & $+$0.03  & $+$0.17 & $-$0.09 & $+$0.04 & $+$0.04 & 0.20 & 0.10 \\
Ce\,{\sc ii}   & $+$0.01  & $+$0.17 & $-$0.08 & $+$0.03 & $+$0.05 & 0.20 & 0.12 \\
Nd\,{\sc ii}   & $+$0.03  & $+$0.18 & $-$0.06 & $+$0.05 & $+$0.06 & 0.21 & 0.18 \\
\hline
\end{tabular}
\end{table*}

\par  We also estimated the influence of model errors, such as uncertainties 
in the effective temperatures and surface gravities, on the derived CNO
abundances. Uncertainties in the carbon abundances result in variation of nitrogen 
abundances, since the CN molecular lines were used for the N abundance determination.  
The variations of the abundance due to changes in effective temperature ($+$140~K), 
surface gravity ($+$0.4~dex), and CNO abundances are summarized in Table~10 for 
HD 87109.  In the last column we present the resulting abundance uncertainties $\sigma_{\rm tot}$
calculated as the square root of the squares of the various sources of uncertainty. 
Derived CNO abundances are weakly sensitive to the variations of the microturbulent velocity 
since weak lines were used for their determination. Calculations of the carbon isotopic ratios 
do not depend on the uncertainties in the C and N abundances and molecular parameters. 
The errors in the $^{12}$C/$^{13}$C determinations are mainly due to uncertainties in the observed 
spectra, such as possible contamination by unidentified atomic or molecular lines, or uncertainties in
the continuum placement.

\begin{table*} 
\centering
\caption{Effect of errors in atmospheric parameters and carbon, oxygen, and nitrogen abundances 
on the CNO abundances.}
\begin{tabular}{cccccccc}\hline\hline
Species & $\Delta T_{\rm eff}$ & $\Delta\log g$ & $\Delta\xi$ & $\Delta\log {\rm (C)}$ 
& $\Delta\log{\rm (N)}$ & $\Delta\log{\rm (O)}$& $\sigma_{\rm tot}$ \\ 
& $+$140~K       &    $+$0.4    & $+$0.3~km\,s$^{-1}$& $+$0.20  & $+$0.20   & $+$0.20   & \\
\hline 
C       & $+$0.08        & $+$0.02      & 0.00      & --          & $-$0.01   & $+$0.05  & 0.10 \\
N       & $+$0.22        & $+$0.05      & $\!\! -$0.01 & $-$0.25  & ---       & $+$0.10  & 0.35\\
O       & $+$0.03        & $+$0.18      & 0.00      & $-$0.01     & $-$0.01   &  ---    & 0.18 \\ 
\hline
\end{tabular}
\end{table*}

\section{Discussion}

\subsection{The rotational velocity}

\par As was presented in Table~1, four giant stars of NGC~3114 
(HD~87109, HD~87479, HD~87833, and HD~304864) have rotational velocities
higher than 2~km\,s$^{-1}$, which is a typical value found among the
cool giant stars (Carlberg at al. 2011).  Of about 1300 stars
investigated by these authors, 30\% have rotational velocities between
5 and 10~km\,s$^{-1}$ and only 2\% of the sample (24 stars) have
rotational velocities higher than 10~km\,s$^{-1}$.  In Fig. 4 we
show the spectra of HD 87566, HD 87833, HD 304864, and HD 87479 where
we can see the effects of the rotation on the broadening of stellar
absorption lines compared to the low rotating giant HD~87566.
Figure~5 shows the positions of HD~87479 and HD~304864 in the $v\sin
i$ $versus$ photometric temperature diagram of Carlberg et al. (2011).
Therefore, the discovery of HD~87479 as a rapid rotating giant star
raises interesting questions. One possibility is that HD~87479 is in
reality a binary system and is forced to co-rotate with its companion,
and so described as a tidally locked binary. In this case,
HD~87479 would be a double-lined SB2 spectroscopic binary. However, our
data show that it has a single-line spectrum.  If HD~87479 is a single
star, its rapid rotation could be explained as a consequence of dregde-up of
angular momentum from a fast rotating core or of the accretion of a planet 
(Carlberg et al. 2011).

\par In open clusters rapid rotating giants are very rare. In
M\,67, Melo et al. (2001) showed that the fast rotators are turn-off
stars, and that the rotational velocity declines along the subgiant
and red giant branches. Other studies found a similar behavior for
dwarfs and giants in open clusters (Pace et al. 2010).  However,
in the open cluster Berkeley~21, one giant (T\,406) is a fast rotator
and also a lithium-rich star (Young et al. 2005). This giant rotates
with a speed similar to that of HD~87479. Planet accretion is the most
likely explanation to account for the lithium enrichment and the
rapid rotation for this star.  It is interesting to note that another
giant star in this cluster, T\,33, analyzed by Hill \& Pasquini
(1999), is also a lithium-rich star but is not a fast rotator.
The stars HD~87479 and HD~304864 set another constraint for possible internal or
external mechanisms in order to explain their high rotational
velocities and no lithium enrichment.

\begin{figure}
\centering
\includegraphics[width=9.5cm]{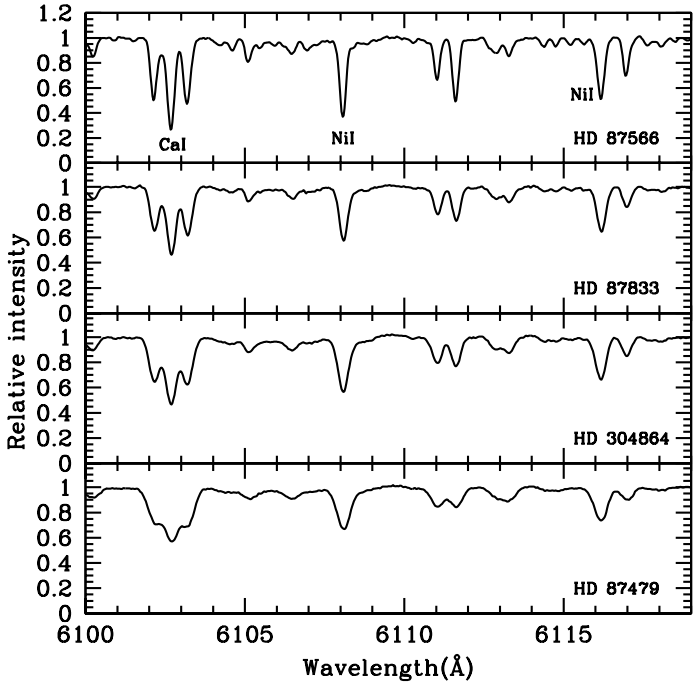}
\caption{Normalized spectra of HD 87566, HD 87833, HD 304864, and HD 87479 
with some identified absorption lines. HD 87566 has a low rotational velocity 
$v\sin i$ $<$ 4.5 km\,s$^{-1}$. In the spectra of HD~87833, HD~304864, and HD~87479  
we see the absorption lines broadened by the rotation velocity of 8.0 km\,s$^{-1}$, 
11 km\,s$^{-1}$, and 15 km\,s$^{-1}$.}
\end{figure}

\begin{figure}
\centering
\includegraphics[width=9.5cm]{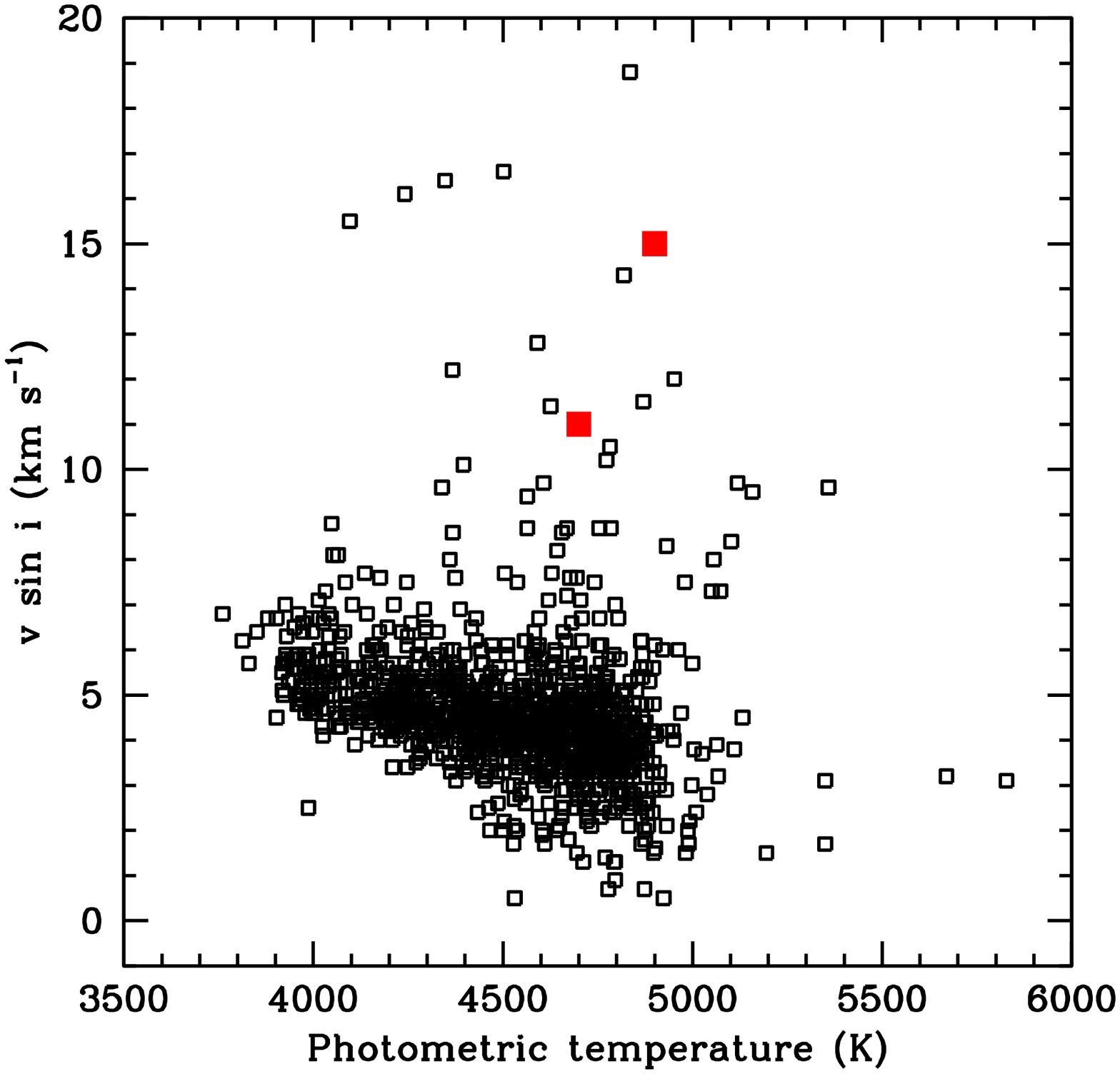}
\caption{Projected rotational velocities for giant stars as a function of photometric
temperature. The open squares represent the stars from the sample studied by Carlberg et al.
(2011) and the red squares the stars HD~87479 and HD~304864.}
\end{figure}

\subsection{The Abundance pattern}

\subsubsection{Metallicity}

\par The open cluster NGC~3114 has solar metallicity. Clari\'a et al. (1989) derived a
mean photometric metallicity of [Fe/H]\,$=\!-0.04$ and Twarog et al. (1997) derived a 
photometric metallicity of [Fe/H]\,=\,+0.02. Excluding HD~87526, our mean cluster 
metallicity is $-$0.01$\pm$0.03 which is in agreement with these previous studies.

\par As shown in Table~4, HD~87526 is a metal-poor star, suggesting
that this star does not belong to the cluster. In fact, Frye et al. (1970) 
using $UBV$ photometry concluded that HD~87526 is a foreground star. 
The results for the abundance ratios of the $\alpha$-elements given in Table~7 
provides further evidence that HD~87526 may belong to the thick disk/halo.  
The [$\alpha$/Fe] ratio is 0.21$\pm$0.27 as given by mean abudance of the elements 
Mg, Si, Ca, and Ti which is a typical value for the stars in this metallicity
range (Carretta et al.  2002). Its abundance pattern is also discussed together with 
the cluster giants. In the following sections we discuss the abundance pattern of the
stars analyzed in this work.

\subsubsection{Carbon, nitrogen, and oxygen}

\par The cluster giants analyzed in this work display [C/Fe] and
     [N/Fe] ratios similar to the giants of the same metallicity
     studied by Mishenina et al. (2006) and Luck \& Heiter (2007).
     The [C/Fe], [N/Fe], and [O/Fe] ratios, from the two papers 
     mentioned above, were computed using absolute abundances obtained in
     these papers and then converted to [X/Fe] ratios using the solar
     abundances adopted in this work (Sect.~3.3).  As shown in
     Fig.~6, our [C/Fe] and [N/Fe] ratios for the cluster giants of
     NGC~3114 have similar values to the giants analyzed by Mishenina
     et al. (2006) and Luck \& Heiter (2007).

\par In dwarf stars there is no trend for the [N/Fe] ratio $versus$
     [Fe/H] in the metallicity range $-2.0<$\,[Fe/H]\,$<$\,+0.3, in 
     other words, the [N/Fe] ratio is $\approx 0.0$ (Clegg et al. 1981; Tomkin
     \& Lambert 1984; Carbon et al.  1987). However, as a star becomes
     giant, because of the deepening of its convective envelope, nuclear
     processed material is brought from the interior to the outer
     layers of the star changing the surface composition. As a
     consequence of the first dredge-up process, the abundance of
     $^{12}$C is reduced and the abundance of $^{14}$N is enhanced
     (Lambert 1981).  Our results for the [C/Fe] and [N/Fe] show the
     effects of the first dredge-up; carbon is underabundant (all
     giants in NGC~3114 have [C/Fe]$<$0.0) while nitrogen is
     overabundant.

\par Carbon abundance in HD~87526 follows the corresponding
value for the stars of the same metallicity in the Galaxy (Fig.~6). 
HD~87526 has a high [N/Fe] ratio of +1.15 and HD~167768, analyzed
by Luck \& Heiter (2007), also displays a high [N/Fe] ratio of +0.91, at a
metallicity of [Fe/H] = --0.61, similar to the metallicity of HD~87526.

\par Figure~6 also shows the abundances of carbon and nitrogen of the
giants of NGC 3114 compared with other cluster giants.  There are
not many available CN abundance determinations for red giants of open
clusters, yet we found some results from the analysis of NGC~6067 and
IC~4725 (Luck 1994), IC~4651 (Mikolaitis et al. 2011a, 2011b),
NGC~7789 (Tautvai$\check{\rm s}$ien\.e et al. 2005) and IC~4756,
NGC~2360, NGC~2447, NGC~3532, NGC~5822, NGC~6134, NGC~6281, and
NGC~6633 (Smiljanic et al. 2009).  The [N/C] ratio has been used as a
diagnostic of the first dredge-up in cluster giants (Smiljanic et
al. 2009) and also for comparison with the evolutionary models of
Schaller et al. (1992). We found a mean value of 0.64$\pm$0.11 for the
cluster giants of NGC~3114 which is in a good agreement with the value
0.6 predicted by Schaller et al. (1992) for a star of mass
$4\,M_{\odot}$.  For the ratio [N/C] we did not find any noticeable
difference between the giants in NGC~3114 and the giants of other
clusters.  Figure 6 shows the oxygen abundance in our cluster
  giants compared with the giants of the same metallicity analyzed by
  Mishenina et al. (2006) and Luck \& Heiter (2007).  We see that the
  giants of NGC~3114 have lower [O/Fe] ratios than the giants of these
  two samples. This figure also shows other [O/Fe] determinations for
  old and young clusters.

\begin{figure}
\includegraphics[height=11cm,width=9.6cm]{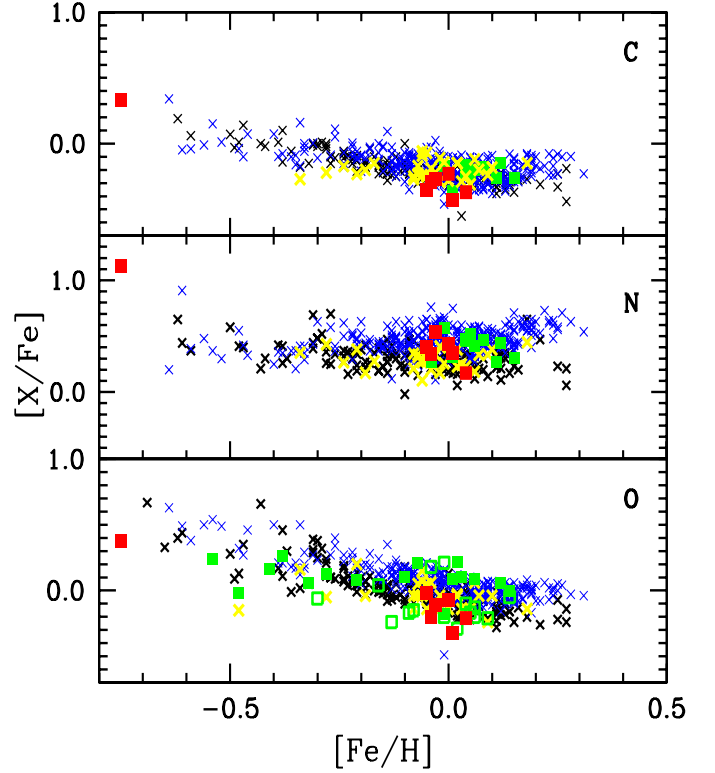}
\caption{Abundance ratios [X/Fe] $versus$ [Fe/H]. {\sl Red squares}:
  cluster giants of NGC 3114; {\sl blue crosses}: field giants of Luck
  \& Heiter (2007); {\sl black crosses}: clump giants of Mishenina et
  al. (2006); {\sl yellow crosses}: clump giants of Tautvai$\check{\rm
    s}$ien\.e et al. (2010); {\sl green filled squares}: mean
  abundances of other open clusters (with ages higher than 1.0 Gyr);
  {\sl green open squares}: clusters with ages less than 1.0
  Gyr. Data from young clusters were taken from Smiljanic et
al. (2009) (IC 4756, NGC 3532, NGC 2447, NGC 2360, NGC 6281, NGC
6633); Pancino et al. 2010 (NGC 2099); Jacobson et al. (2011) (NGC
1245 \& NGC 1817); Carrera \& Pancino (2010) (Hyades \& Praesepe);
Za$\check{\rm c}$s et al. (2011) (NGC 1545 \& Tr 2); Villanova et
al. (2009) (NGC~6475); Brown et al. (1996) (Mellote 71). Data from old
clusters were taken from Pacino et al. (2010) (NGC~2420, NGC 7789, Cr
100, M\,67); Carrera \& Pancino (2011) (Berkeley 32 \& NGC 752);
Jacobson et al. (2011) (NGC 2158); Jacobson et al. (2008) (NGC 7142);
Friel et al. (2010) (NGC 188 \& Berkeley 39); Carreta et al. (2005)
(Cr 261); Gratton \& Contarini (1994) (Mellote 66 \& NGC 2243); Friel
et al. (2005) (Berkeley 17); Young et al. (2005) (Berkeley 20 \&
Berkeley 29).}
\end{figure}

\subsubsection{$^{12}$C/$^{13}$C ratio}

\par The carbon isotopic ratio has already been investigated in some
open clusters, nonetheless there are few papers dedicated to obtaining
this ratio.  Previous carbon isotopic ratio determinations in open
clusters were done by Gilroy (1989), Luck (1994), Tautvai$\check{\rm
  s}$ien\.e et al. (2000), Smiljanic et al. (2009), and Mikolaitis et
al. (2010, 2011a, 2011b). Figure~7 shows the carbon isotopic
  ratio for stars of NGC~3114 analyzed in this paper and of other
  clusters as a function of the cluster turn-off mass.  In this
figure, we also show the expected carbon isotopic ratio from the
standard first dredge-up models of Schaller et al. (1992). Gilroy
(1989) was the first to show that there is a change in the carbon
isotopic ratio for clusters with turn-off mass less than $2.2
M_{\odot}$. Smiljanic et al. (2009) were able to show that some
cluster giants with turn-off masses higher than $2.0 M_{\odot}$ could
also have low carbon isotopic ratios as the low-mass stars.  These low
ratios seen in more massive stars are currently attributed to
non-canonical mixing or thermohaline convection (Stancliffe et
al. 2007).

\par We inspected whether there would be any connection between the
position of a giant in the color-magnitude diagram (Fig.~1) and
$^{12}$C/$^{13}$C\, and [N/C] ratios.  The stars 6 (=\,HD~87109),
150 (=\,HD~87479), and 283 (=\,HD 304864) are probably post-helium
flash giants, and the stars 181 (=\,HD~87566), 238 (=\,HD~304859), and
262 (=\,HD~87833) seem to be first-ascent giants. We did not detect
any significant variations in the carbon isotopic ratio among these
classes of giants. As far as the [N/Fe] ratio is concerned, only the
star 238 (=\,HD~304859) has a relatively low ratio.

\par The open cluster analyzed in this work adds more data points
  at one turn-off mass in Fig.~7 so that a determination of the
  $^{12}$C/$^{13}$C ratio in more cluster giants would be very
  important for understanding the physics of the dredge-up and mixing
  phenomena.

\begin{figure}
\includegraphics[height=11cm,width=9.6cm]{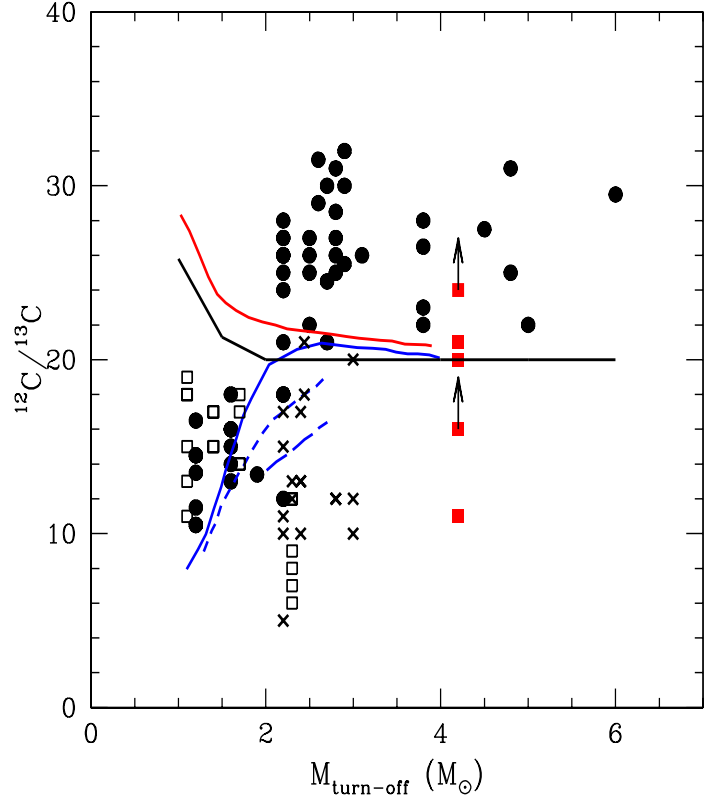}
\caption{The $^{12}$C/$^{13}$C ratio in giants of open clusters
  $versus$ the turn-off mass. Data points represent the isotopic ratio
  for each star of the cluster. Filled circles are data from Gilroy
  (1989), open squares data from Smiljanic et al. (2009), and crosses
  data from Mikolaitis et al. (2011a, 2011b, 2012). Red squares
  represent the stars of NGC~3114. The theoretical $^{12}$C/$^{13}$C
  ratio as a function of stellar mass is taken from the models of
  Schaller et al. (1992), {\sl black solid line}.  Other models from
  Charbonnel \& Lagarde (2010) are also shown: standard, {\sl red
    solid line}; thermohaline mixing, {\sl blue solid line};
  thermohaline and rotation-induced mixing with 110 km\,s$^{-1}$, {\sl
    blue short-dashed line}; thermohaline and rotation-induced mixing
  with 250 km\,s$^{-1}$, {\sl blue long-dashed line.}}
\end{figure}

\subsubsection{Li}

\par In their thorough study of the Li abundance in G-K giants, Brown
et al. (1989) showed that field giants have a mean lithium abundance
of $\log\varepsilon$(Li) $\approx$\,+0.1.  In giants of open clusters,
lithium abundances have already been investigated by Gilroy (1989),
Luck (1994), Pasquini et al. (2001) for NGC~3680, Pasquini et
al. (2004) for IC~4651, Gonzalez \& Wallerstein (2000) for M\,11 and
Za$\check{\rm c}$s et al. (2011) for NGC~1545 and Tr\,2. The main
result of these investigations showed that lithium abundance in
evolved K giants is very low ($-\!1.0\! <\! \log\varepsilon({\rm
  Li})\! <\! 1.0$), lower than predicted by the standard evolution
models of Iben (1966, 1967). An extramixing mechanism is the most
likely explanation of the low lithium abundances observed in cluster
giants (Pasquini et al. 2001; Gilroy 1989).

\par The mean value of lithium abundance for the giants of NGC~3114 is
$\log\varepsilon$(Li)\,=\,0.9$\pm$0.4 and is slighty above the value
for cluster giants with a turn-off mass of $4.0\,M_{\odot}$ analyzed
by Gilroy (1989). Lithium abundance was derived for the young open
cluster M\,11 by Gonzalez \& Wallerstein (2000). For this cluster
having the same turn-off mass $4\,M_{\odot}$ as NGC~3114, they found
a mean lithium abundance of 1.2$\pm$0.3.

\subsubsection{Other elements: Na to Ni}

\par Sodium overabundance has been observed in the atmospheres
of A-F supergiant stars (Denisenkov \& Ivanov 1987). According to
these authors, sodium is synthesized in the convective core of
main-sequence stars in the NeNa cycle.  Mixing at the first dredge-up
carries products of the CNO cycle to the surface of stars.  Therefore,
one should expect sodium enrichment in supergiants and giants but not 
in dwarfs. In fact, Fig.~2 of Boyarchuk et al. (2001) shows
that [Na/Fe] is anti-correlated with $\log g$.  It is known that sodium
lines suffer from NLTE effects which lead to an overestimation of the
sodium abundances. To account for NLTE effects we used the theoretical
work of Gratton et al. (1999) who calculated the values of the NLTE
corrections for several sodium lines using a grid of different
atmospheric parameters and equivalent widths of the sodium lines. For
HD~87526, a non-member of the cluster, we found a correction of
$\approx$0.3 dex for the lines 6154.22\,\AA\, and 6160.75\,\AA\, used
for the sodium abundance determination considering the temperature and
gravity of this star. Therefore, for HD 87526 we have
[Na/Fe]\,=\,+0.36. For the giants of NGC~3114 with equivalent widths
of the analyzed sodium lines around 120\,m\AA, the NLTE corrections
are less than 0.2~dex.

\par For aluminum, the ratio [Al/Fe] observed in our cluster giants
has the same value as the field giants of the same metallicity.
In HD~87526 this ratio is in agreement with the trend seen for the
stars of similar metallicity.

\par As for the $\alpha$-elements defined in Sect.~4.2.1, our cluster
giants follow the same trend as seen in the local disk field giants
studied by Luck \& Heiter (2007) and disk red-clump giants studied by
Mishenina et al. (2006).

\par Iron peak elements are formed in large amounts in Type Ia
supernovae and all these elements should follow the same trend with
iron abundance. Indeed, nickel does remain constant with
[Ni/Fe]\,=\,0.0 for $-$0.8 $<$ [Fe/H] $<$ 0.1 (Luck \& Heiter
2007). For the element chromium, the study for the local giants of
Luck \& Heiter (2007) found a tight correlation between [Fe/H] and
[Cr/H] ratios with a correlation coefficient of 1.06, i.e.,
[Cr/Fe]\,$\simeq$\,0.0 for $-$0.8 $<$ [Fe/H] $<$ 0.2, despite some
discrepant points as they reported. Our cluster giant ratios also
follow the general trend seen for the field giants.

\par In HD~87526 the abundance ratios [X/Fe] of iron group elements
(Cr and Ni) have similar values as dwarf stars of the same
metallicity (Edvardsson et al. 1993).

\subsubsection{Heavy elements: neutron-capture elements}

\par In field giant stars, abundances of s-process elements were
investigated by Mishenina et al. (2006, 2007) and Luck \& Heiter
(2007).  In both of these studies the authors found that the abundance
ratios [X/Fe] for Y, La, Ce, and Nd remain flat at 0.0 between $-$0.7
$<$ [Fe/H] $<$+0.3. Figure 8 shows these ratios for individual
neutron-capture elements as well as the {\sl mean} values of
these s-elements, in the notation [s/Fe].

\par The s-process element abundances observed in HD~87526 also follow
the same trend as in giants and dwarfs of the same metallicity.  For
yttrium and neodymium we also used the results of Fulbright (2000) and
Edvardsson et al. (1993) for the dwarf stars with metallicities between
$-$1.0 $<$ [Fe/H] $<$ $-$0.5 for comparison with our results for HD~87526.

\begin{figure}
\centering
\includegraphics[height=11cm,width=9.6cm]{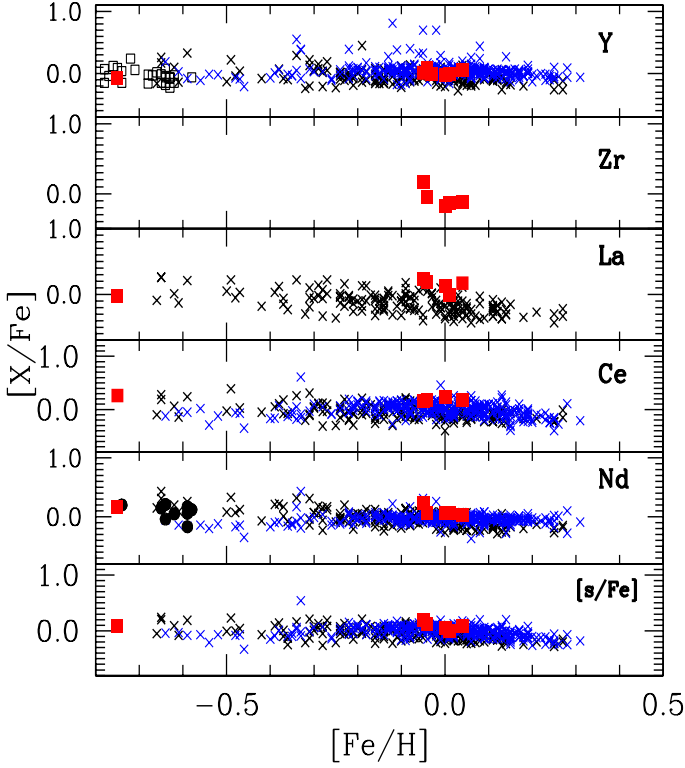}
\caption{Abundance ratios [X/Fe] $versus$ [Fe/H]. Symbols have the
  same meaning as in Fig.~6. In addition we provide results for
  yttrium (open squares; Fulbright 2000) and neodymium (filled
  circles, Edvardsson et al. 1993) for dwarf stars to compare with the
  abundances of HD~87526.}
\end{figure}

\par We also comment on the behavior of the heavy-element abundance
pattern with age.  The studies of heavy-element abundance pattern in
cluster giants are scarce in the literature and are mostly based on
one or two elements with only a few lines analyzed. Zirconium is one
of these elements with abundances based on few lines (Maiorca et
al. 2011, see also Friel et al. 2010). In our sample we were able to
measure more than three lines of zirconium in the spectra of all stars
except the metal-poor star HD 87526 and the high-rotating stars
HD~87479 and HD~304864 (Table~1). Because of high rotation some weaker
lines in the spectra of these stars become washed away in rotationally
broadened profiles. For this reason we were able to measure only a few
lines of zirconium in the spectrum of HD~304864 and none in the
spectrum of HD~87479.

\par Recently, Maiorca et al. (2011) presented the results of their
investigation on the abundance pattern of s-process elements based on
a large sample of 19 open clusters. In this study, after analyzing a
sample of dwarfs and giants in open clusters, they noticed that the
youngest open clusters were the most s-process enriched. In Fig.~9
we show the mean abundance of s-process elements using the results
presented in Maiorca et al. (2011) and the mean s-process abundance
from Table 8 of this work. The open cluster NGC~3114 also follows the trend 
noticed by Maiorca et al. (2011).

\begin{figure}
\includegraphics[height=11cm,width=9.6cm]{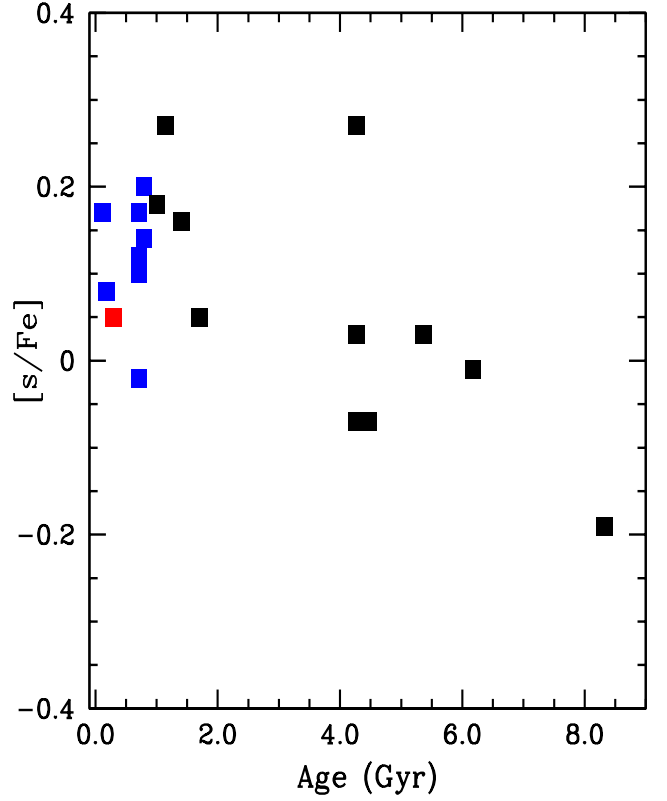}
\caption{{\sl Mean} s-process abundances $versus$ age for NGC~3114 compared
with a sample of the open clusters analyzed by Mairoca et al. (2001).}
\end{figure}

\section{Conclusions}

\par The main conclusions of our abundance analysis employing
high-resolution optical spectra of the giants in NGC~3114 can be
summarized as follows:

\begin{enumerate}

\item NGC~3114 is a young cluster (0.16~Gyr) with a turn-off mass of
  $4.2\,M_{\odot}$.  It has a solar metallicity of
  [Fe/H]\,=\,$-$0.01$\pm$0.03 in agreement with previous photometric
  studies.  There are a few clusters with turn-off mass determined,
  therefore the determination of the turn-off mass for NGC~3114 adds
  another point to this small sample.

\item The star HD 87526 belongs to the thick disk/Galactic halo
  because the metallicity and the abundances are very different from
  the other giants of the cluster and is not a member of NGC~3114.

\item Two stars have high rotational velocities higher than the other stars in the
clusters. HD 87479 and HD 304864 have 15.0 km\,s$^{-1}$ and 11.0 km\,s$^{-1}$.

\item Analysis of the light elements reveals that carbon has a low
  abundance and that nitrogen shows an enrichment, similar to field
  giant stars and giants of other clusters.  The [N/C] ratio is in
  good agreement with the predictions given by the models of Schaller
  et al. (1992) after the first dredge-up. The carbon isotopic ratio
  (Fig.~5) lies slighty below the predicted value for a cluster with
  a turn-off mass of $4.0\,M_{\odot}$. The derived value of
  $^{12}$C/$^{13}$C\,$\geq$18 for NGC~3114 is similar to the value for
  M\,11 (16 -- 20), a cluster with the same age and turn-off mass
  (Gonzalez \& Wallerstein 2000).

\item The open cluster NGC~3114 follows the trend of heavy-element abundance 
seen in other open clusters (Maiorca et al. 2011),  that is, an enrichment of 
s-process elements in young clusters compared to old clusters.

\item Further spectroscopic analysis of other young clusters will be
  welcome in order to investigate the abundance pattern in
  intermediate mass stars as they evolve away from the
  main-sequence. On the other hand, some studies for dwarfs in young
  clusters (Ford et al. 2005; St\"utz et al. 2006; Pace et al. 2010;
  Villanova et al. 2009) showed that oxygen has solar abundance and
  shows no depletion seen in giants of these clusters. This
  indicates that evolutionary aspects and mixing changed the abundance
  of clusters giants.  Further abundance analysis of dwarfs and (a
  few) subgiant stars in these clusters will also be welcome.

\end{enumerate}

\begin{acknowledgements} N.A.D acknowledges the support of the PCI/MCTI (Brazil) grant under the 
Project 311.868/2011-8. This research has made use of the SIMBAD 
database, operated at CDS, Strasbourg, France.
\end{acknowledgements}

{}

\Online

\longtab{2}{
\begin{longtable}{lcccccccccc}
\caption{Observed Fe\,{\sc i} and Fe\,{\sc ii} lines}\\
\hline\hline
\multicolumn{4}{c}{} & \multicolumn{7}{c}{Equivalent widths (m\AA)}\\
\hline
\multicolumn{4}{c}{} & \multicolumn{7}{c}{HD}\\
\hline
Element & $\lambda$ & $\chi$(eV) & $\log gf$ & 87109 & 87479 & 87526 & 87566 & 87833 & 304859 & 304864 \\\hline
Fe\,{\sc i}\,& 5198.71 &   2.22 &   $-$2.140 & --- & --- &  136 & --- &  --- &    --- &    --- \\
& 5242.49 &   3.63 &   $-$0.970 &    --- &    --- &    109 &    149 &    141 &    144 &    --- \\
& 5250.21 &   0.12 &   $-$4.920 &    --- &    --- &    112 &    --- &    --- &    --- &    --- \\
& 5253.03 &   2.28 &   $-$3.790 &     90 &    --- &    --- &     99 &     84 &     98 &    --- \\
& 5288.52 &   3.69 &   $-$1.510 &    --- &    --- &     67 &    --- &    124 &    134 &    --- \\
& 5307.36 &   1.61 &   $-$2.970 &    --- &    --- &    132 &    --- &    --- &    --- &    --- \\
& 5315.05 &   4.37 &   $-$1.400 &     97 &    --- &     22 &     98 &     91 &     99 &    105 \\
& 5321.11 &   4.43 &   $-$1.190 &     92 &    110 &     42 &     88 &     90 &     85 &    --- \\
& 5322.04 &   2.28 &   $-$2.840 &    --- &    --- &     80 &    142 &    139 &    --- &    --- \\
& 5364.87 &   4.45 &      0.230 &    --- &    --- &    140 &    --- &    --- &    --- &    --- \\
& 5373.71 &   4.47 &   $-$0.710 &    115 &    118 &     64 &     99 &     92 &    106 &    127 \\
& 5389.48 &   4.42 &   $-$0.250 &    --- &    --- &    --- &    135 &    130 &    127 &    --- \\
& 5400.50 &   4.37 &   $-$0.100 &    --- &    --- &    135 &    --- &    --- &    --- &    --- \\
& 5417.03 &   4.42 &   $-$1.530 &     83 &     88 &     26 &     77 &     72 &     84 &     85 \\
& 5441.34 &   4.31 &   $-$1.580 &     79 &     74 &     28 &     77 &     72 &     83 &     90 \\
& 5445.04 &   4.39 &      0.041 &    --- &    --- &    129 &    149 &    --- &    --- &    --- \\
& 5522.45 &   4.21 &   $-$1.400 &     95 &     98 &     38 &     88 &     93 &     88 &    108 \\
& 5531.98 &   4.91 &   $-$1.460 &    --- &     71 &    --- &    --- &    --- &     70 &    --- \\
& 5532.75 &   3.57 &   $-$2.000 &    --- &    --- &     51 &    --- &    --- &    --- &    --- \\
& 5554.90 &   4.55 &   $-$0.380 &    --- &    --- &     87 &    --- &    --- &    --- &    --- \\
& 5560.21 &   4.43 &   $-$1.040 &     96 &    106 &     44 &     90 &     81 &     87 &    110 \\
& 5563.60 &   4.19 &   $-$0.840 &    --- &    --- &     93 &    --- &    --- &    109 &    --- \\
& 5567.39 &   2.61 &   $-$2.560 &    --- &    --- &     69 &    --- &    --- &    --- &    --- \\
& 5584.77 &   3.57 &   $-$2.170 &    --- &    --- &    --- &     95 &    --- &    106 &    --- \\
& 5624.02 &   4.39 &   $-$1.330 &    --- &    --- &    --- &    100 &    --- &    105 &    --- \\
& 5633.95 &   4.99 &   $-$0.120 &    123 &    120 &     61 &    106 &    101 &    110 &    --- \\
& 5635.82 &   4.26 &   $-$1.740 &     78 &     91 &     21 &     75 &     66 &     75 &    --- \\
& 5638.26 &   4.22 &   $-$0.720 &    139 &    --- &     73 &    133 &    116 &    137 &    --- \\
& 5686.53 &   4.55 &   $-$0.450 &    --- &    --- &     84 &    121 &    --- &    136 &    --- \\
& 5691.50 &   4.30 &   $-$1.370 &    104 &    111 &     33 &    --- &     92 &    --- &    116 \\
& 5705.47 &   4.30 &   $-$1.360 &     87 &    --- &     24 &     83 &     70 &     83 &     89 \\
& 5717.83 &   4.28 &   $-$0.979 &    --- &    --- &    --- &    126 &    --- &    --- &    --- \\
& 5731.76 &   4.26 &   $-$1.150 &    119 &    --- &     48 &    111 &    102 &    108 &    --- \\
& 5762.99 &   4.21 &   $-$0.410 &    --- &    --- &    119 &    --- &    --- &    --- &    --- \\
& 5806.73 &   4.61 &   $-$0.900 &    104 &    106 &     46 &     99 &     88 &     99 &    123 \\
& 5814.81 &   4.28 &   $-$1.820 &     59 &    --- &    --- &     62 &     54 &     68 &     78 \\
& 5852.22 &   4.55 &   $-$1.180 &    100 &    105 &     25 &    --- &     89 &    --- &    111 \\
& 5883.82 &   3.96 &   $-$1.210 &    --- &    135 &     66 &    128 &    115 &    115 &    --- \\
& 5934.65 &   3.93 &   $-$1.020 &    --- &    --- &     80 &    140 &    117 &    139 &    --- \\
& 6024.06 &   4.55 &   $-$0.060 &    --- &    --- &    121 &    --- &    --- &    --- &    --- \\
& 6027.05 &   4.08 &   $-$1.090 &    134 &    144 &     68 &    109 &    108 &    118 &    --- \\
& 6056.01 &   4.73 &   $-$0.400 &    118 &    --- &     65 &    108 &     97 &    109 &    125 \\
& 6079.01 &   4.65 &   $-$0.970 &     97 &    113 &    --- &     92 &     84 &     89 &     94 \\
& 6082.71 &   2.22 &   $-$3.580 &    128 &    --- &    --- &    --- &    --- &    --- &    --- \\
& 6093.64 &   4.61 &   $-$1.350 &     83 &    --- &    --- &     75 &     73 &     75 &     82 \\
& 6096.66 &   3.98 &   $-$1.780 &     97 &     95 &     23 &     82 &     79 &     85 &     96 \\
& 6105.13 &   4.55 &   $-$2.050 &     37 &     59 &    --- &    --- &    --- &    --- &    --- \\
& 6120.25 &   0.91 &   $-$5.950 &     72 &     65 &    --- &     84 &    --- &     85 &     81 \\
& 6151.62 &   2.18 &   $-$3.290 &    148 &    135 &     52 &    138 &    114 &    138 &    --- \\
& 6157.73 &   4.08 &   $-$1.110 &    --- &    142 &     70 &    --- &    134 &    --- &    --- \\
& 6165.36 &   4.14 &   $-$1.470 &    103 &    105 &     31 &     93 &     79 &     98 &    109 \\
& 6170.51 &   4.79 &   $-$0.380 &    --- &    --- &     70 &    --- &    --- &    --- &    --- \\
& 6173.34 &   2.22 &   $-$2.880 &    --- &    --- &     92 &    --- &    --- &    --- &    --- \\
& 6187.99 &   3.94 &   $-$1.570 &    112 &    110 &     37 &    101 &     90 &    109 &    117 \\
& 6200.31 &   2.60 &   $-$2.440 &    --- &    --- &     84 &    --- &    142 &    --- &    --- \\
& 6213.43 &   2.22 &   $-$2.480 &    --- &    --- &    111 &    --- &    --- &    --- &    --- \\
& 6265.13 &   2.18 &   $-$2.550 &    --- &    --- &    127 &    --- &    --- &    --- &    --- \\
& 6322.69 &   2.59 &   $-$2.430 &    --- &    --- &     92 &    --- &    --- &    --- &    --- \\
& 6380.74 &   4.19 &   $-$1.320 &    --- &    117 &     50 &    --- &    116 &    --- &    --- \\
& 6392.54 &   2.28 &   $-$4.030 &     89 &    --- &    --- &     94 &    --- &    --- &    --- \\
& 6411.65 &   3.65 &   $-$0.660 &    --- &    --- &    140 &    --- &    --- &    --- &    --- \\
& 6436.41 &   4.19 &   $-$2.460 &     52 &     61 &    --- &     59 &    --- &     57 &    --- \\
& 6469.19 &   4.83 &   $-$0.620 &    --- &    122 &     49 &    --- &    109 &    --- &    135 \\
& 6551.68 &   0.99 &   $-$5.790 &     97 &     70 &    --- &    --- &    --- &    --- &     92 \\
& 6574.23 &   0.99 &   $-$5.020 &    --- &    --- &    --- &    145 &    --- &    --- &    --- \\
& 6591.31 &   4.59 &   $-$2.070 &    --- &     59 &    --- &     50 &    --- &    --- &    --- \\
& 6593.87 &   2.44 &   $-$2.420 &    --- &    --- &    111 &    --- &    --- &    --- &    --- \\
& 6597.56 &   4.79 &   $-$0.920 &     85 &     94 &    --- &     85 &     83 &     84 &     90 \\
& 6608.03 &   2.28 &   $-$4.030 &    100 &    --- &    --- &    102 &    --- &    100 &    109 \\
& 6609.11 &   2.56 &   $-$2.690 &    --- &    --- &     77 &    --- &    --- &    --- &    --- \\
& 6646.93 &   2.61 &   $-$3.990 &     80 &     90 &    --- &    --- &    --- &    --- &     97 \\
& 6653.85 &   4.14 &   $-$2.520 &     44 &    --- &    --- &     40 &     38 &     47 &    --- \\
& 6699.14 &   4.59 &   $-$2.190 &     36 &    --- &    --- &     42 &    --- &     41 &     43 \\
& 6703.57 &   2.76 &   $-$3.160 &    132 &    123 &    --- &    125 &    --- &    132 &    139 \\
& 6704.48 &   4.22 &   $-$2.660 &     26 &    --- &    --- &     31 &    --- &     32 &    --- \\
& 6713.74 &   4.79 &   $-$1.600 &     62 &     43 &    --- &     57 &     42 &     58 &     57 \\
& 6739.52 &   1.56 &   $-$4.950 &     86 &     84 &    --- &     93 &    --- &     94 &     81 \\
& 6745.96 &   4.07 &   $-$2.770 &     21 &    --- &    --- &     36 &    --- &     32 &     35 \\
& 6750.15 &   2.42 &   $-$2.620 &    --- &    --- &     96 &    --- &    --- &    --- &    --- \\
& 6752.71 &   4.64 &   $-$1.200 &    --- &    109 &     27 &    --- &    --- &    --- &    --- \\
& 6783.70 &   2.59 &   $-$3.980 &    --- &    --- &    --- &    --- &    --- &    100 &    --- \\
& 6793.26 &   4.07 &   $-$2.470 &     48 &    --- &    --- &     47 &    --- &     56 &    --- \\
& 6806.85 &   2.73 &   $-$3.210 &    123 &    114 &    --- &    124 &    --- &    118 &    134 \\
& 6810.26 &   4.61 &   $-$0.990 &    101 &    105 &     33 &     86 &     84 &     95 &    103 \\
& 6820.37 &   4.64 &   $-$1.170 &    103 &     89 &     30 &     94 &     79 &     96 &    103 \\
& 6841.34 &   4.61 &   $-$0.600 &    --- &    --- &     70 &    --- &    130 &    --- &    --- \\
& 6851.64 &   1.61 &   $-$5.320 &     77 &     68 &    --- &    --- &    --- &    --- &     84 \\
& 6858.15 &   4.61 &   $-$0.930 &     99 &    117 &    --- &     88 &     98 &     95 &    129 \\
& 7132.99 &   4.08 &   $-$1.610 &    --- &    109 &    --- &    109 &    --- &    108 &    116 \\
& 7540.43 &   2.73 &   $-$3.850 &    --- &    --- &    --- &     76 &    --- &    --- &    --- \\\hline
Fe\,{\sc ii}\,& 4993.35 &   2.81 &   $-$3.670 &  96 & --- &  --- &  75 &  78 &   73 &    --- \\
& 5132.66 &   2.81 &   $-$4.000 &     93 &    --- &    --- &    --- &    --- &    --- &    --- \\
& 5234.62 &   3.22 &   $-$2.240 &    --- &    --- &    --- &    111 &    129 &    120 &    --- \\
& 5325.56 &   3.22 &   $-$3.170 &     99 &    135 &    111 &     65 &    --- &     71 &    --- \\
& 5414.05 &   3.22 &   $-$3.620 &     83 &    --- &    --- &     62 &     66 &     58 &    --- \\
& 5425.25 &   3.20 &   $-$3.210 &    102 &    118 &    108 &     73 &     77 &     71 &     92 \\
& 5991.37 &   3.15 &   $-$3.560 &    109 &    111 &     88 &     71 &     87 &     63 &    109 \\
& 6084.10 &   3.20 &   $-$3.800 &     85 &     92 &     69 &     56 &     60 &     50 &    --- \\
& 6149.25 &   3.89 &   $-$2.720 &     97 &    115 &     92 &     59 &     74 &    --- &    111 \\
& 6247.55 &   3.89 &   $-$2.340 &    127 &    133 &    --- &     75 &     98 &     70 &    127 \\
& 6416.92 &   3.89 &   $-$2.680 &     83 &    --- &     94 &     64 &     64 &     67 &     89 \\
& 6432.68 &   2.89 &   $-$3.580 &     99 &    113 &     92 &     70 &     79 &     79 &    104 \\
\hline
\end{longtable}
}

\longtab{5}{
\begin{longtable}{cccccccccccc}
\caption{Other lines studied}\\
\hline\hline
\multicolumn{5}{c}{} &\multicolumn{7}{c}{Equivalent widths (m\AA)} \\\hline
\multicolumn{5}{c}{} & \multicolumn{7}{c}{HD}\\
\hline
Element & $\lambda$ & $\chi$(eV) & $\log gf$ & Ref & 87109 & 87479 & 87526 & 87566 & 87833 & 304859 & 304864 \\
\hline

Na\,{\sc i} &  5682.65 &  2.10 & $-$0.70 &  PS   & --- &  --- &  128 &  --- &  --- &  --- &  --- \\
Na\,{\sc i} &  6154.22 &  2.10 & $-$1.51 &  PS   & 118 &  107 &   51 &  120 &  105 &  118 &  118 \\
Na\,{\sc i} &  6160.75 &  2.10 & $-$1.21 &  R03  & 136 &  --- &   74 &  134 &  126 &  132 &  129 \\

Mg\,{\sc i} &  4730.04 &  4.34 &  $-$2.39 & R03   & --- & ---  & --- & 115 &  --- & --- & --- \\ 
Mg\,{\sc i} &  5711.10 &  4.34 &  $-$1.75 & R99   & --- & ---  & 107 & --- &  132 & --- & --- \\
Mg\,{\sc i} &  6318.71 &  5.11 &  $-$1.94 & Ca07  &  91 & ---  & --- &  97 &   81 & 106 & --- \\
Mg\,{\sc i} &  6319.24 &  5.11 &  $-$2.16 & Ca07  &  55 & ---  & --- &  56 &   53 &  73 & --- \\
Mg\,{\sc i} &  6319.49 &  5.11 &  $-$2.67 & Ca07  &  24 & ---  & --- &  36 &   19 &  28 & --- \\
Mg\,{\sc i} &  6765.45 &  5.75 &  $-$1.94 & MR94  & --- & ---  & --- & --- &   43 & --- & --- \\
Mg\,{\sc i} &  7387.70 &  5.75 &  $-$0.87 & MR94  &  90 & ---  & --- &  92 &   85 &  92 & --- \\
Mg\,{\sc i} &  8712.69 &  5.93 &  $-$1.26 & WSM   &  69 & ---  & --- & --- &   69 & --- & --- \\
Mg\,{\sc i} &  8717.83 &  5.91 &  $-$0.70 & WSM   & --- & ---  &  76 & 114 &  116 & 129 & --- \\
Mg\,{\sc i} &  8736.04 &  5.94 & $-$0.34  & WSM   & --- & ---  & 118 & 138 &  144 & --- & --- \\

Al\,{\sc i} & 6696.03 & 3.14 & $-$1.48 & MR94& --- & --- &  --- &  81 & --- &  88 & --- \\
Al\,{\sc i} & 6698.67 & 3.14 & $-$1.63 & R03 &  64 &  76 &  --- &  67 &  52 &  68 & --- \\
Al\,{\sc i} & 7835.32 & 4.04 & $-$0.58 & R03 &  63 & 117 &  23  & --- &  82 &  79 & --- \\
Al\,{\sc i} & 7836.13 & 4.02 & $-$0.40 & R03 &  83 &  71 &  26  &  91 &  82 &  93 & --- \\
Al\,{\sc i} & 8772.88 & 4.02 & $-$0.25 & R03 & 120 & --- &  56  &  98 &  95 & 102 & --- \\
Al\,{\sc i} & 8773.91 & 4.02 & $-$0.07 & R03 & --- & 135 & 100  & 113 & 141 & 127 & --- \\

Si\,{\sc i} & 5793.08 & 4.93 & $-$2.06 &  R03 &  94 & --- &  49 & 76 &  77 &  75 &  98 \\
Si\,{\sc i} & 6125.03 & 5.61 & $-$1.54 &  E93 &  70 & --- &  23 & 45 &  47 &  61 &  75 \\
Si\,{\sc i} & 6131.58 & 5.62 & $-$1.68 &  E93 &  47 & --- & --- & 40 & --- &  45 & --- \\
Si\,{\sc i} & 6145.02 & 5.61 & $-$1.43 &  E93 &  60 & --- &  34 & 52 &  55 &  56 &  75 \\
Si\,{\sc i} & 6155.14 & 5.62 & $-$0.77 &  E93 & 112 & 112 &  65 & 99 &  92 & 105 & 109 \\
Si\,{\sc i} & 7760.64 & 6.20 & $-$1.28 &  E93 &  29 &  54 &  14 & 27 & --- &  32 &  46 \\
Si\,{\sc i} & 7800.00 & 6.18 & $-$0.72 &  E93 &  78 & --- &  47 & 62 &  67 & --- & --- \\
Si\,{\sc i} & 8728.01 & 6.18 & $-$0.36 &  E93 &  99 & --- &  97 & 75 &  95 &  89 & 139 \\
Si\,{\sc i} & 8742.45 & 5.87 & $-$0.51 &  E93 & 113 & --- &  82 & 86 & --- & 110 & 138 \\

Ca\,{\sc i} & 6161.30 & 2.52 & $-$1.27 &   E93 &  134 & --- &  52 &  138 &  115 & 145 &  --- \\
Ca\,{\sc i} & 6166.44 & 2.52 & $-$1.14 &   R03 &  131 & 146 &  50 &  130 &  112 & 130 &  143 \\
Ca\,{\sc i} & 6169.04 & 2.52 & $-$0.80 &   R03 &  --- & --- &  87 &  --- &  139 & --- &  --- \\
Ca\,{\sc i} & 6169.56 & 2.53 & $-$0.48 &  DS91 &  --- & --- & 113 &  --- &  --- & --- &  --- \\
Ca\,{\sc i} & 6455.60 & 2.51 & $-$1.29 &   R03 &  124 & 117 &  51 &  128 &  104 & 130 &  125 \\
Ca\,{\sc i} & 6471.66 & 2.51 & $-$0.69 &   S86 &  --- & --- & 104 &  --- &  130 & --- &  --- \\
Ca\,{\sc i} & 6493.79 & 2.52 & $-$0.11 &  DS91 &  --- & --- & 143 &  --- &  --- & --- &  --- \\

Ti\,{\sc i} & 4512.74 & 0.84 & $-$0.480 & MFK & --- & --- &  68 & --- & --- & --- & --- \\
Ti\,{\sc i} & 4562.64 & 0.02 & $-$2.660 & MFK &  98 & --- & --- & 112 & --- & 110 & --- \\
Ti\,{\sc i} & 4617.28 & 1.75 & $+$0.389 & MFK & 144 & --- &  58 & 143 & 125 & 143 & 141 \\
Ti\,{\sc i} & 4758.12 & 2.25 & $+$0.420 & MFK & --- & --- &  29 & 106 & --- & 109 & 122 \\
Ti\,{\sc i} & 4759.28 & 2.25 & $+$0.511 & MFK & 107 & --- &  25 & 116 &  96 & 118 & --- \\
Ti\,{\sc i} & 4778.26 & 2.24 & $-$0.330 & MFK &  59 &  58 & --- &  75 &  61 &  80 &  61 \\
Ti\,{\sc i} & 5016.17 & 0.85 & $-$0.570 & MFK & --- & --- & --- & --- & 137 & --- & --- \\
Ti\,{\sc i} & 4820.41 & 1.50 & $-$0.440 & MFK & --- & --- & --- & --- & --- & --- & 149 \\
Ti\,{\sc i} & 5039.96 & 0.02 & $-$1.130 & MFK & --- & --- &  93 & --- & --- & --- & --- \\
Ti\,{\sc i} & 5043.59 & 0.84 & $-$1.730 & MFK & 112 & 103 & --- & 115 & --- & 112 & --- \\
Ti\,{\sc i} & 5062.10 & 2.16 & $-$0.460 & MFK & --- & --- & --- &  78 &  60 & --- & --- \\
Ti\,{\sc i} & 5113.45 & 1.44 & $-$0.780 & E93 &  94 & --- & --- & --- & --- & 106 & --- \\
Ti\,{\sc i} & 5145.47 & 1.46 & $-$0.570 & MFK & 114 & --- &  24 & 121 & --- & 123 & --- \\
Ti\,{\sc i} & 5173.75 & 0.00 & $-$1.120 & MFK & --- & --- & 107 & --- & --- & --- & --- \\
Ti\,{\sc i} & 5210.39 & 0.05 & $-$0.879 & MFK & --- & --- & 106 & --- & --- & --- & --- \\
Ti\,{\sc i} & 5219.71 & 0.02 & $-$2.290 & MFK & 138 & --- &  21 & --- & 116 & --- & 134 \\
Ti\,{\sc i} & 5223.63 & 2.09 & $-$0.561 & MFK & --- & --- & --- &  82 & --- &  92 & --- \\
Ti\,{\sc i} & 5282.44 & 1.05 & $-$1.300 & MFK & --- & --- & --- & --- &  74 & --- & --- \\
Ti\,{\sc i} & 5295.78 & 1.05 & $-$1.631 & MFK &  70 & --- & --- &  91 &  66 &  94 &  88 \\
Ti\,{\sc i} & 5490.16 & 1.46 & $-$0.932 & MFK & --- & --- & --- & --- & --- & 116 & --- \\
Ti\,{\sc i} & 5662.16 & 2.32 & $-$0.110 & MFK & --- & --- & --- & --- &  89 & --- & --- \\
Ti\,{\sc i} & 5689.48 & 2.30 & $-$0.470 & MFK &  60 &  56 & --- &  79 &  58 &  80 &  57 \\
Ti\,{\sc i} & 5866.46 & 1.07 & $-$0.839 & E93 & --- & --- &  39 & --- & 139 & --- & --- \\
Ti\,{\sc i} & 5922.12 & 1.05 & $-$1.470 & MFK & --- & 117 & --- & 117 & 100 & 116 & 124 \\
Ti\,{\sc i} & 5978.55 & 1.87 & $-$0.500 & MFK &  93 &  93 & --- & 105 &  82 & 115 & --- \\
Ti\,{\sc i} & 6091.18 & 2.27 & $-$0.420 & R03 &  75 &  67 & --- &  88 &  68 &  88 &  73 \\
Ti\,{\sc i} & 6126.22 & 1.07 & $-$1.420 & R03 & 118 & 112 &  16 & 128 & 101 & 132 & 123 \\
Ti\,{\sc i} & 6258.11 & 1.44 & $-$0.360 & MFK & --- & --- &  41 & --- & --- & --- & --- \\
Ti\,{\sc i} & 6261.11 & 1.43 & $-$0.480 & B86 & --- & --- &  36 & --- & 143 & --- & --- \\
Ti\,{\sc i} & 6554.24 & 1.44 & $-$1.220 & MFK & 106 &  83 & --- & 113 &  85 & --- & --- \\

Cr\,{\sc i} & 4801.03 & 3.12 & $-$0.130 & MFK & --- & --- &  49 &  121 & 105 & 118 & 114 \\
Cr\,{\sc i} & 4814.26 & 3.09 & $-$1.170 & MFK &  63 & --- & --- &   75 & --- & --- & --- \\
Cr\,{\sc i} & 4836.85 & 3.10 & $-$1.140 & MFK &  73 & --- & --- &  --- &  60 &  70 & --- \\
Cr\,{\sc i} & 4936.34 & 3.11 & $-$0.220 & MFK &  88 & --- &  45 &   92 & --- & 109 & --- \\
Cr\,{\sc i} & 4954.80 & 3.12 & $-$0.140 & MFK & --- & --- & --- &   97 & --- & 111 & --- \\
Cr\,{\sc i} & 5193.50 & 3.42 & $-$0.900 & MFK &  60 & --- & --- &   53 & --- &  52 & --- \\
Cr\,{\sc i} & 5200.18 & 3.38 & $-$0.530 & MFK &  82 & --- & --- &   84 & --- &  88 & --- \\
Cr\,{\sc i} & 5214.13 & 3.37 & $-$0.740 & MFK &  45 & --- & --- &   54 &  46 &  52 & --- \\
Cr\,{\sc i} & 5214.61 & 3.32 & $-$0.660 & MFK &  79 & --- & --- &   88 &  66 &  89 &  66 \\
Cr\,{\sc i} & 5238.96 & 2.71 & $-$1.300 & MFK &  67 & --- & --- &  --- &  46 &  75 & --- \\
Cr\,{\sc i} & 5272.00 & 3.45 & $-$0.420 & MFK &  52 & --- & --- &  --- & --- & --- & --- \\
Cr\,{\sc i} & 5296.70 & 0.98 & $-$1.240 & GS  & --- & --- & 127 &  --- & --- & --- & --- \\
Cr\,{\sc i} & 5300.75 & 0.98 & $-$2.130 & GS  & 149 & --- & --- &  --- & 138 & --- & --- \\
Cr\,{\sc i} & 5304.18 & 3.46 & $-$0.690 & MFK &  41 & --- & --- &   49 &  62 &  73 &  65 \\
Cr\,{\sc i} & 5312.86 & 3.45 & $-$0.561 & MFK &  81 & --- & --- &   79 &  65 &  60 &  75 \\
Cr\,{\sc i} & 5318.77 & 3.44 & $-$0.690 & MFK &  47 & --- & --- &   61 &  47 &  65 & --- \\
Cr\,{\sc i} & 5348.32 & 1.00 & $-$1.290 & GS  & --- & --- & 131 &  --- & --- & --- & --- \\
Cr\,{\sc i} & 5628.65 & 3.42 & $-$0.770 & MFK &  36 & --- & --- &   45 &  29 &  52 & --- \\
Cr\,{\sc i} & 5702.32 & 3.45 & $-$0.680 & MFK &  45 & --- & --- &  --- & --- & --- & --- \\
Cr\,{\sc i} & 5781.18 & 3.32 & $-$0.879 & MFK &  50 & --- & --- &  --- &  37 & --- & --- \\
Cr\,{\sc i} & 5781.75 & 3.32 & $-$0.750 & MFK &  82 & --- & --- &   84 &  64 &  84 & --- \\
Cr\,{\sc i} & 5783.07 & 3.32 & $-$0.400 & MFK & --- &  73 &  20 &   83 &  73 &  87 &  90 \\
Cr\,{\sc i} & 5783.86 & 3.32 & $-$0.300 & GS  & --- & 100 &  32 &  --- & --- & --- & 123 \\
Cr\,{\sc i} & 5784.97 & 3.32 & $-$0.380 & MFK &  78 & --- & --- &   88 &  61 &  91 & --- \\
Cr\,{\sc i} & 5787.92 & 3.32 & $-$0.080 & GS  &  96 & 100 &  41 &   97 &  84 &  99 & 102 \\
Cr\,{\sc i} & 6330.09 & 0.94 & $-$2.870 & R03& 139 & 106 &  26 &  133 & 100 & 134 & 128 \\

Ni\,{\sc i} & 4904.42 & 3.54 & $-$0.190 & MFK &  --- & ---  &  95 & --- & --- & --- & --- \\
Ni\,{\sc i} & 4913.98 & 3.74 & $-$0.600 & MFK &  108 & ---  & --- &  96 & 110 &  98 & 101 \\
Ni\,{\sc i} & 4935.83 & 3.94 & $-$0.340 & MFK &  132 & 104  &  49 & --- & --- & --- & 101 \\
Ni\,{\sc i} & 4953.21 & 3.74 & $-$0.620 & MFK &  125 & 111  &  46 & --- & 101 & --- & 114 \\
Ni\,{\sc i} & 4967.52 & 3.80 & $-$1.600 & MFK &   35 & ---  & --- & --- &  46 & --- & --- \\
Ni\,{\sc i} & 4995.66 & 3.63 & $-$1.611 & MFK &   66 & ---  & --- &  73 & --- &  70 & --- \\
Ni\,{\sc i} & 5003.75 & 1.68 & $-$3.130 & MFK &  --- & ---  & --- & 122 & --- & --- & --- \\
Ni\,{\sc i} & 5010.94 & 3.63 & $-$0.900 & MFK &  --- &  86  &  36 &  85 &  85 & --- & 104 \\
Ni\,{\sc i} & 5084.11 & 3.68 & $-$0.180 & E93 &  --- & ---  & --- & --- & --- & --- & 136 \\
Ni\,{\sc i} & 5094.42 & 3.83 & $-$1.120 & MFK &   68 &  64  & --- &  68 & --- &  65 &  62 \\
Ni\,{\sc i} & 5115.40 & 3.83 & $-$0.280 & R03 &  --- & ---  & --- & --- & --- & --- & 138 \\
Ni\,{\sc i} & 5157.98 & 3.61 & $-$1.719 & MFK &   49 & ---  & --- & --- & --- & --- & --- \\
Ni\,{\sc i} & 5197.17 & 3.90 & $-$1.140 & MFK &   75 & ---  & --- &  81 &  64 &  80 & --- \\
Ni\,{\sc i} & 5578.73 & 1.68 & $-$2.670 & MFK &  144 & ---  &  62 & --- & 135 & --- & --- \\
Ni\,{\sc i} & 5587.87 & 1.94 & $-$2.370 & MFK &  --- & ---  & --- & 135 & --- & 143 & --- \\
Ni\,{\sc i} & 5589.37 & 3.90 & $-$1.150 & MFK &   64 & ---  & --- & --- & --- &  65 &  58 \\
Ni\,{\sc i} & 5593.75 & 3.90 & $-$0.790 & MFK &   85 &  83  &  28 &  81 & --- &  81 &  90 \\
Ni\,{\sc i} & 5643.09 & 4.17 & $-$1.250 & MFK &   40 & ---  & --- &  37 &  27 &  47 & --- \\
Ni\,{\sc i} & 5748.36 & 1.68 & $-$3.250 & MFK &  120 & ---  & --- & 124 & 103 & 121 & --- \\
Ni\,{\sc i} & 5760.84 & 4.11 & $-$0.810 & MFK &   89 & ---  & --- &  90 &  76 &  94 & --- \\
Ni\,{\sc i} & 5805.23 & 4.17 & $-$0.600 & MFK &   79 &  70  &  24 &  75 &  64 &  76 &  70 \\
Ni\,{\sc i} & 5847.01 & 1.68 & $-$3.440 & MFK &  108 & 105  &  16 & 114 &  92 & 114 & 114 \\
Ni\,{\sc i} & 5996.74 & 4.24 & $-$1.060 & MFK &   50 &  42  &  14 &  53 &  46 &  50 &  53 \\
Ni\,{\sc i} & 6053.69 & 4.24 & $-$1.070 & MFK &   72 & ---  & --- & --- &  64 &  65 & --- \\
Ni\,{\sc i} & 6086.29 & 4.27 & $-$0.470 & MFK &   77 &  81  & --- &  72 & --- &  80 &  82 \\
Ni\,{\sc i} & 6108.12 & 1.68 & $-$2.489 & MFK &  --- & ---  &  69 & --- & 137 & --- & --- \\
Ni\,{\sc i} & 6111.08 & 4.09 & $-$0.830 & MFK &   72 &  78  &  23 &  75 &  69 &  75 &  91 \\
Ni\,{\sc i} & 6128.98 & 1.68 & $-$3.390 & MFK &  123 & ---  &  28 & 118 & 100 & 122 & 125 \\
Ni\,{\sc i} & 6130.14 & 4.27 & $-$0.979 & MFK &   55 & ---  &  13 &  53 &  49 &  61 &  64 \\
Ni\,{\sc i} & 6176.82 & 4.09 & $-$0.260 & R03 &  120 & ---  & --- & 100 &  99 & 104 & 128 \\
Ni\,{\sc i} & 6177.25 & 1.83 & $-$3.600 & MFK &   84 & ---  & --- &  74 &  72 &  82 &  78 \\
Ni\,{\sc i} & 6186.72 & 4.11 & $-$0.900 & MFK &   67 & ---  & --- &  65 &  63 &  75 &  77 \\
Ni\,{\sc i} & 6204.61 & 4.09 & $-$1.150 & MFK &   68 & ---  & --- &  77 &  54 &  67 &  72 \\
Ni\,{\sc i} & 6223.99 & 4.11 & $-$0.971 & MFK &   76 & ---  &  19 &  81 &  71 &  68 &  83 \\
Ni\,{\sc i} & 6230.10 & 4.11 & $-$1.200 & MFK &  --- & ---  & --- & --- &  55 & --- & --- \\
Ni\,{\sc i} & 6322.17 & 4.15 & $-$1.210 & MFK &   56 & ---  & --- &  57 &  48 &  61 & --- \\
Ni\,{\sc i} & 6327.60 & 1.68 & $-$3.090 & MFK &  144 & 124  &  40 & --- & 116 & 139 & 137 \\
Ni\,{\sc i} & 6378.26 & 4.15 & $-$0.821 & MFK &   86 &  84  &  28 &  86 &  72 &  84 &  88 \\
Ni\,{\sc i} & 6384.67 & 4.15 & $-$1.000 & MFK & ---  & ---  &  21 &  55 &  75 &  61 & --- \\
Ni\,{\sc i} & 6482.81 & 1.94 & $-$2.851 & MFK &  136 & ---  & --- & 138 & 110 & 139 & --- \\
Ni\,{\sc i} & 6532.88 & 1.94 & $-$3.420 & MFK &  --- & ---  & --- &  81 & --- &  82 & --- \\
Ni\,{\sc i} & 6586.32 & 1.95 & $-$2.790 & MFK &  135 & 135  &  45 & 139 & 110 & 131 & 128 \\
Ni\,{\sc i} & 6598.61 & 4.24 & $-$0.932 & MFK &   61 & ---  & --- &  60 &  55 &  63 & --- \\
Ni\,{\sc i} & 6635.14 & 4.42 & $-$0.750 & MFK &   56 &  55  & --- &  73 &  54 &  73 & --- \\
Ni\,{\sc i} & 6767.78 & 1.83 & $-$2.110 & MFK &  --- & ---  & 101 & --- & --- & --- & --- \\
Ni\,{\sc i} & 6772.32 & 3.66 & $-$1.010 & R03 &  131 & 121  &  43 & 120 & 104 & 111 & 127 \\
Ni\,{\sc i} & 6842.04 & 3.66 & $-$1.440 & E93 &   87 & ---  & --- &  90 &  76 &  89 &  88 \\

Y\,{\sc ii} & 5087.43 & 1.08 & $-$0.170 & SN96 & --- & --- & 128 & 110 & --- & 112 & --- \\
Y\,{\sc ii} & 5200.41 & 0.99 & $-$0.570 & SN96 & --- & --- & 127 & 113 &  98 & 118 & --- \\
Y\,{\sc ii} & 5205.72 & 1.03 & $-$0.340 & SN96 & --- & --- & 103 & --- & --- & --- & --- \\
Y\,{\sc ii} & 5289.81 & 1.03 & $-$1.850 & VWR  &  63 &  43 &  22 &  49 &  37 &  53 & 146 \\
Y\,{\sc ii} & 5402.78 & 1.84 & $-$0.440 & R03  &  84 &  74 &  51 &  62 &  59 &  67 &  91 \\

Zr\,{\sc i} & 4772.30 & 0.62 & $-$0.060 & A05  & --- & --- & --- &  63 &  42 &  61 & --- \\ 
Zr\,{\sc i} & 4805.87 & 0.69 & $-$0.580 & A05  &  18 & --- & --- &  33 &  18 &  41 & --- \\
Zr\,{\sc i} & 4815.05 & 0.65 & $-$0.380 & A05  & --- & --- & --- &  54 & --- & --- & --- \\
Zr\,{\sc i} & 4828.05 & 0.62 & $-$0.750 & A05  &  10 & --- & --- &  26 & --- &  27 & --- \\
Zr\,{\sc i} & 5385.13 & 0.52 & $-$0.640 & A05  &  31 & --- & --- &  46 &  27 &  45 &  25 \\
Zr\,{\sc i} & 5620.13 & 0.52 & $-$1.090 & A05  & --- & --- & --- & --- & --- &  43 & --- \\
Zr\,{\sc i} & 5955.34 & 0.00 & $-$1.699 & A05  & --- & --- & --- & --- & --- &  25 & --- \\
Zr\,{\sc i} & 6032.60 & 1.48 & $-$0.350 & A05  & --- & --- & --- & --- & --- &  17 & --- \\
Zr\,{\sc i} & 6127.46 & 0.15 & $-$1.060 & S96  &  43 & --- & --- &  56 &  36 &  70 &  36 \\
Zr\,{\sc i} & 6134.57 & 0.00 & $-$1.280 & S96  &  34 & --- & --- &  54 &  25 &  65 & --- \\
Zr\,{\sc i} & 6140.46 & 0.52 & $-$1.410 & S96  &  12 & --- & --- &  14 &  13 &  24 & --- \\
Zr\,{\sc i} & 6143.18 & 0.07 & $-$1.100 & S96  &  55 & --- & --- &  70 &  32 &  72 &  53 \\

La\,{\sc ii} & 4086.71 & 0.00 & $-$0.160 & SN96 & --- & --- & 100 & --- & --- &  --- & --- \\
La\,{\sc ii} & 4934.83 & 1.25 & $-$0.920 & VWR  & --- & --- & --- & --- & --- &  --- & --- \\
La\,{\sc ii} & 5303.53 & 0.32 & $-$1.350 & VWR  &  65 & --- &  24 &  52 &  43 &   53 & --- \\
La\,{\sc ii} & 5880.63 & 0.24 & $-$1.830 & VWR  & --- & --- & --- & --- & --- &   37 & --- \\
La\,{\sc ii} & 6320.43 & 0.17 & $-$1.520 & SN96 &  79 & --- &  14 &  63 &  47 &   70 & --- \\
La\,{\sc ii} & 6390.48 & 0.32 & $-$1.410 & VWR  &  83 & --- &  26 & --- & --- &   75 &  77 \\
La\,{\sc ii} & 6774.33 & 0.12 & $-$1.700 & VWR  &  68 & --- & --- & --- & --- &   57 &  53 \\

Ce\,{\sc ii} & 4073.47 & 0.48 & $+$0.320 & SN96  & --- & --- &  74 & --- & --- & --- & --- \\
Ce\,{\sc ii} & 4083.23 & 0.70 & $+$0.240 & SN96  & --- & --- &  69 & --- & --- & --- & --- \\
Ce\,{\sc ii} & 4120.84 & 0.32 & $-$0.240 & SN96  & --- & --- &  49 & --- & --- & --- & --- \\
Ce\,{\sc ii} & 4222.60 & 0.12 & $-$0.180 & SN96  & --- & --- & 103 & --- & --- & --- & --- \\
Ce\,{\sc ii} & 4418.79 & 0.86 & $+$0.310 & SN96  & --- & --- &  73 & --- & --- & --- & --- \\
Ce\,{\sc ii} & 4486.91 & 0.30 & $-$0.360 & SN96  & --- & --- &  66 & --- & --- & --- & --- \\
Ce\,{\sc ii} & 4562.37 & 0.48 & $+$0.330 & SN96  & --- & --- &  95 & 103 &  90 & 101 & --- \\
Ce\,{\sc ii} & 4628.16 & 0.52 & $+$0.260 & SN96  & --- & --- &  90 & --- &  80 & --- & --- \\
Ce\,{\sc ii} & 5117.17 & 1.40 & $+$0.01  & VWR   &  43 & --- &  16 &  31 & --- &  29 & --- \\
Ce\,{\sc ii} & 5187.46 & 1.21 & $+$0.30  & VWR   &  67 & --- &  33 &  59 &  49 &  56 & --- \\
Ce\,{\sc ii} & 5274.24 & 1.28 & $+$0.38  & VWR   &  80 & --- &  39 &  62 & --- &  60 & --- \\
Ce\,{\sc ii} & 5330.58 & 0.87 & $+$0.28  & VWR   & --- & --- & --- & --- & --- & --- & --- \\
Ce\,{\sc ii} & 5472.30 & 1.25 & $-$0.19  & VWR   &  54 & --- &  20 &  41 & --- &  38 & --- \\
Ce\,{\sc ii} & 6051.80 & 0.23 & $-$1.60  & S96   &  30 & --- & --- &  26 & --- &  30 & --- \\

Nd\,{\sc ii} & 4811.34 & 0.06 & $-$1.015 & VWR  & --- & --- &  52 & --- &  76 &  85 & --- \\
Nd\,{\sc ii} & 4959.12 & 0.06 & $-$0.916 & VWR  & --- & --- &  56 & --- & --- & --- & 117 \\
Nd\,{\sc ii} & 4989.95 & 0.63 & $-$0.624 & VWR  & --- & --- &  32 & --- &  57 & --- &  79 \\
Nd\,{\sc ii} & 5063.72 & 0.98 & $-$0.758 & VWR  &  39 & --- & --- & --- &  19 & --- & --- \\
Nd\,{\sc ii} & 5092.80 & 0.38 & $-$0.510 & E93  &  92 & --- & --- &  78 & --- &  73 & --- \\
Nd\,{\sc ii} & 5130.59 & 1.30 & $+$0.100 & SN96 & --- & --- &  46 & --- & --- & --- & --- \\
Nd\,{\sc ii} & 5212.36 & 0.20 & $-$0.700 & SN96 & --- & --- & --- & --- & --- & --- & 111 \\
Nd\,{\sc ii} & 5234.19 & 0.55 & $-$0.460 & SN96 & --- & --- &  39 & --- &  67 & --- & --- \\
Nd\,{\sc ii} & 5293.16 & 0.82 & $-$0.200 & SN96 & --- & --- &  68 & --- & --- & --- & --- \\
Nd\,{\sc ii} & 5311.46 & 0.98 & $-$0.560 & SN96 & --- & --- &  16 & --- & --- & --- & --- \\
Nd\,{\sc ii} & 5319.81 & 0.55 & $-$0.350 & SN96 & --- & --- &  68 & --- &  78 &  89 & --- \\
Nd\,{\sc ii} & 5361.47 & 0.68 & $-$0.400 & SN96 & --- & --- &  60 & --- & --- & --- & --- \\
Nd\,{\sc ii} & 5416.38 & 0.86 & $-$0.980 & VWR  &  32 & --- & --- &  34 &  17 &  21 & --- \\
Nd\,{\sc ii} & 5431.54 & 1.12 & $-$0.457 & VWR  &  51 & --- &  14 & --- &  26 & --- &  55 \\
Nd\,{\sc ii} & 5442.26 & 0.68 & $-$0.900 & SN96 &  73 & --- & --- & --- &  44 & --- &  65 \\
Nd\,{\sc ii} & 5740.88 & 1.16 & $-$0.560 & VWR  &  46 & --- & --- &  39 &  24 &  33 & --- \\
Nd\,{\sc ii} & 5842.39 & 1.28 & $-$0.601 & VWR  &  29 & --- & --- &  21 &  16 &  18 &  26 \\
\hline
\footnote{References to Table 4:
\par A05:  Antipova et al. (2005);   
\par B86:  Blackwell D.E. et al. (1986);   
\par Ca07: Carretta et al. (2007);    
\par DS91: Drake \& Smith (1991);    
\par E93:  Edvardsson et al. (1993);    
\par GS:   Gratton \& Sneden (1988);   
\par MR94: Mcwilliam \& Rich (1994);   
\par MFW:  Martin et al. (1988);   
\par PS:   Preston \& Sneden (2001);   
\par R03:  Reddy et al. (2003);   
\par R04:  Reyniers et al. (2004);    
\par R99:  Reddy et al. (1999);   
\par S86:  Smith et al. (1986);    
\par S96:  Smith et al. (1996);    
\par SN96: Sneden et al. (1996);    
\par VWR:  van Wickel \& Reyners (2000);   
\par WSM:  Wiese, Smith \& Miles (1969). 
}\\
\end{longtable}
}

\end{document}